# ORTHOGONAL MATRIX IN CRYPTOGRAPHY

Yeray Cachón Santana

Member of CriptoRed (U.P.M.)

**ABSTRACT**

**In this work is proposed a method using orthogonal matrix transform properties to encrypt and decrypt a message. It will be showed how to use matrix's functions to create complex encryptions.**

**Because orthogonal matrix are always diagonalizable on R, and the exponential of a diagonal matrix is easy to compute, the exponential of orthogonal matrix will be used to encrypt text messages.**

**INTRODUCTION**

In several works about cryptography, matrix is used as an application that a vector of plaintext is transform in a vector of ciphertext and, to decrypt, the reverse of the application is used. So, giving a message **M** and a encoding matrix A, the encrypted message is giving by **X**=**A**\***M**. To decrypt, the reversal matrix is used, so the original message is giving by **M**=**A**$^{-1}$\***X**. Other works described the application of Quadratic forms because the inverse of nonsingular matrices of higher order is much difficult [2].

Orthogonal matrices have several properties that make them interesting to diagonalize and find its reverse. This paper proposes a method to encrypt and decrypt a message using the properties of these matrices. Some of these properties will be used and are quicker than other matrix method to decrypt:

- Fast inversion
- Real eigenvalues
- Orthogonal eigenvector

**METHOD**

Let be **A** an orthogonal matrix. Because **A** is orthogonal, **A** is diagonalizable, and can be written as a product of matrix, one of them a diagonal matrix.

It will be showed how to encrypt using exponential of matrix, so firstly, let's see how to compute the exponential of a matrix. So, **A**=**C**$^{-1}$**DC**, where **C** and **C**$^{-1}$ are the change of base matrix and **D** is the diagonal matrix. Let's be $\lambda_i$ the eigenvalues of **A**, and **v**$_i$ the eigenvectors associated. So, to calculate the exponential of **A**:

$$e^{A} = C^{-1}e^{D}C = C^{-1}\begin{pmatrix} e^{\lambda_1} & & & \\ & e^{\lambda_2} & & \\ & & \ddots & \\ & & & e^{\lambda_n} \end{pmatrix}C$$

Let's see how to calculate the exponential of a matrix A that is diagonalizable:

$$e^{A} = \sum_{n} \frac{A^n}{n!} = \sum_{n} \frac{(C^{-1}DC)^n}{n!}$$

$$= \sum_{n} \frac{C^{-1}DCC^{-1}\cdots CC^{-1}DC}{n!}$$

$$= C^{-1}\left(\sum_{n} \frac{D^n}{n!}\right)C = C^{-1}e^{D}C$$

So, the exponential of the matrix D will be:

$$e^D = \sum_n \frac{\mathbf{D}^n}{n!} = \sum_n \frac{1}{n!}\begin{pmatrix} \lambda_1^n & & & & \\ & \lambda_2^n & & & \\ & & \lambda_3^n & & \\ & & & \ddots & \\ & & & & \lambda_n^n \end{pmatrix}$$

$$= \sum_n \begin{pmatrix} \frac{\lambda_1^n}{n!} & & & & \\ & \frac{\lambda_2^n}{n!} & & & \\ & & \frac{\lambda_3^n}{n!} & & \\ & & & \ddots & \\ & & & & \frac{\lambda_n^n}{n!} \end{pmatrix}$$

$$= \begin{pmatrix} e^{\lambda_1} & & & & \\ & e^{\lambda_2} & & & \\ & & e^{\lambda_3} & & \\ & & & \ddots & \\ & & & & e^{\lambda_n} \end{pmatrix}$$

In fact, for any Taylor expandable function, $f(\mathbf{A})$, if $\mathbf{D}$ is the diagonal matrix:

$$f(\mathbf{A}) = \mathbf{C}^{-1} f(\mathbf{D}) \mathbf{C} = \mathbf{C}^{-1}\begin{pmatrix} f(\lambda_1) & & & \\ & f(\lambda_2) & & \\ & & \ddots & \\ & & & f(\lambda_n) \end{pmatrix}\mathbf{C}$$

Because calculate an inversion it takes a long time of computation (if the dimensions of the matrix **A** are large), it can be taken an orthogonal matrix ($\mathbf{C}^{-1}=\mathbf{C}^T$). It will see later than, in fact, having 2 matrices with the public and private keys, if the matrix of public and private keys are orthogonal, their product, will be also orthogonal:

$$\mathbf{C} * \mathbf{C}^T = (\mathbf{k}_{pub}\, \mathbf{k}_{priv})(\mathbf{k}_{pub}\, \mathbf{k}_{priv})^T$$
$$= (\mathbf{k}_{pub}\, \mathbf{k}_{priv})(\mathbf{k}_{priv}^T\, \mathbf{k}_{pub}^T)$$
$$= \mathbf{k}_{pub}(\mathbf{k}_{priv}\, \mathbf{k}_{priv}^T)\mathbf{k}_{pub}^T$$

So, because $k_{pub}$ and $k_{priv}$ are orthogonal, $\mathbf{k}_{priv}\mathbf{k}_{priv}^T = \mathbf{k}_{pub}\mathbf{k}_{pub}^T = \mathbb{I}$, and, then:

$$\mathbf{C} * \mathbf{C}^T = \mathbf{k}_{pub}(\mathbf{k}_{priv}\mathbf{k}_{priv}^T)\mathbf{k}_{pub}^T = \mathbf{k}_{pub} * \mathbf{k}_{pub}^T = \mathbb{I}$$

But, if $\mathbf{C} = \mathbf{k}_{pub}\, \mathbf{k}_{priv}$, the decomposition of an orthogonal matrix in several orthogonal matrix it's not unique. Let's see why:

**Proof**: Let be **A** and **B** orthogonal matrix:

$$\mathbf{A} = \begin{pmatrix} \cos\Theta_1 & -\sin\Theta_1 \\ \sin\Theta_1 & \cos\Theta_1 \end{pmatrix},$$

$$\mathbf{B} = \begin{pmatrix} \cos\Theta_2 & -\sin\Theta_2 \\ \sin\Theta_2 & \cos\Theta_2 \end{pmatrix}$$

Their product **A*B** will also be and orthogonal matrix:

$$\mathbf{A} * \mathbf{B} = \begin{pmatrix} \cos(\Theta_1 + \Theta_2) & -\sin(\Theta_1 + \Theta_2) \\ \sin(\Theta_1 + \Theta_2) & \cos(\Theta_1 + \Theta_2) \end{pmatrix}$$

Being **A*B** orthogonal, any matrix **A** and **B** orthogonal give an **A*B** orthogonal, so the decomposition is not unique.∎

An exponential function is used because is a one-to-one function and it's well-defined in $\Re$ and it's invertible in all $\Re$. So, the one-to-one function is the relationship between the plaintext space (PT) and the ciphertext space (CT):

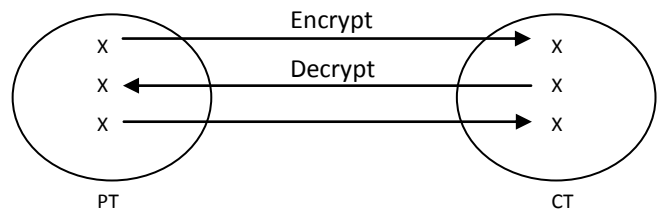

The encryption of the message will be done by blocks, so each plaintext block will be encrypt in a cyphertext block, as it's shown:

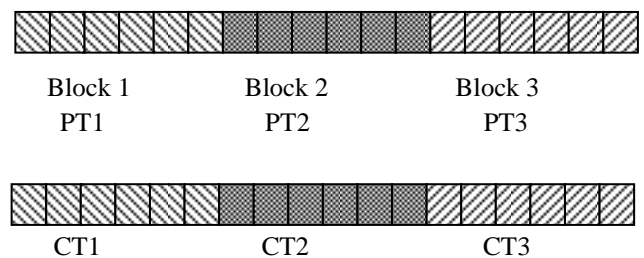

As it will see later, the n-cipher block will be encrypted in a C-space given by $C^n$. Let be **C** the matrix transform from the plain-text to the ciphertext first-space.

$$PT \rightarrow CT^{(1)}$$

Using several times the matrix transform, the ciphertext of the n-th space will be computed:

$$PT \rightarrow CT^{(1)} \rightarrow CT^{(2)} \rightarrow \ldots \rightarrow CT^{(n)}$$

Because after n-th times the n-th power of $C^n=I$, it's not necessary to compute all n-power of C. So, if k is the order of **C**:

$$C^n = C^{k*p+m} = (\mathbb{I})^p * C^n = C^n$$

**APPLICATION**

Let see how to encrypt the word "CRYPTOGRAPHY" choosing blocks of 4 letters. Let **C** the following orthogonal matrix:

$$\mathbf{C} = \frac{1}{\sqrt{2}} \begin{pmatrix} 1 & -1 & 0 & 0 \\ 1 & 1 & 0 & 0 \\ 0 & 0 & 1 & -1 \\ 0 & 0 & 1 & 1 \end{pmatrix}$$

We will use a exponential function as operator because it's a one-to-one function (trigonometric functions aren't one-to-one) and it's well defined in $\mathbb{R}$. Firstly, because it's been chosen blocks of 4 letters (matrix 4x4), the word will be split in blocks of 4 letters:

CRYP | TOGR | APHY

So, to encrypt, we will use as eigenvalues and eigenvector each of the number of alphabet matched with each letter:

C : $\lambda_{1,1} = 3$
R : $\lambda_{1,2} = 18$
Y : $\lambda_{1,3} = 25$
P : $\lambda_{1,4} = 16$

T : $\lambda_{2,1} = 20$
O : $\lambda_{2,2} = 15$
G : $\lambda_{2,3} = 7$
R : $\lambda_{2,4} = 18$

A : $\lambda_{3,1} = 1$
P : $\lambda_{3,2} = 16$
H : $\lambda_{3,3} = 8$
Y : $\lambda_{3,4} = 25$

It will be taken $x_{i,j} = \lambda_{i,j}$.

Once that **C** has been chosen and the eigenvalues and eigenvectors, it will be create the vectors on the space cryptographic.

For each block:

Block 1: "CRYP"

$$\mathbf{Y}_1 = \mathbf{C}^{-1}\mathbf{D}\mathbf{C}\mathbf{X}_1$$
$$= \frac{1}{2}\begin{pmatrix} 1 & 1 & 0 & 0 \\ -1 & 1 & 0 & 0 \\ 0 & 0 & 1 & 1 \\ 0 & 0 & -1 & 1 \end{pmatrix} \begin{pmatrix} e^3 & 0 & 0 & 0 \\ 0 & e^{18} & 0 & 0 \\ 0 & 0 & e^{25} & 0 \\ 0 & 0 & 0 & e^{16} \end{pmatrix} \begin{pmatrix} 1 & -1 & 0 & 0 \\ 1 & 1 & 0 & 0 \\ 0 & 0 & 1 & -1 \\ 0 & 0 & 1 & 1 \end{pmatrix} \begin{pmatrix} 3 \\ 18 \\ 25 \\ 16 \end{pmatrix}$$
$$= \frac{1}{2}\begin{pmatrix} 21e^{18} - 15e^3 \\ 21e^{18} + 15e^3 \\ 41e^{16} + 9e^{25} \\ 41e^{16} - 9e^{25} \end{pmatrix}$$

Block 2: "TOGR"

$$\mathbf{Y}_2 = (\mathbf{C}^2)^{-1}\mathbf{D}\mathbf{C}^2\mathbf{X}_2$$
$$= \begin{pmatrix} 0 & 1 & 0 & 0 \\ -1 & 0 & 0 & 0 \\ 0 & 0 & 0 & 1 \\ 0 & 0 & -1 & 0 \end{pmatrix} \begin{pmatrix} e^{20} & 0 & 0 & 0 \\ 0 & e^{15} & 0 & 0 \\ 0 & 0 & e^7 & 0 \\ 0 & 0 & 0 & e^{18} \end{pmatrix} \begin{pmatrix} 0 & -1 & 0 & 0 \\ 1 & 0 & 0 & 0 \\ 0 & 0 & 0 & -1 \\ 0 & 0 & 1 & 0 \end{pmatrix} \begin{pmatrix} 20 \\ 15 \\ 7 \\ 18 \end{pmatrix}$$
$$= \begin{pmatrix} 20e^{15} \\ 15e^{20} \\ 7e^{18} \\ 18e^7 \end{pmatrix}$$

Block 3: "APHY"

$$\mathbf{Y}_3 = (\mathbf{C}^3)^{-1}\mathbf{D}\mathbf{C}^3\mathbf{X}_3$$
$$= \frac{1}{2}\begin{pmatrix} -1 & 1 & 0 & 0 \\ -1 & -1 & 0 & 0 \\ 0 & 0 & -1 & 1 \\ 0 & 0 & -1 & -1 \end{pmatrix} \begin{pmatrix} e & 0 & 0 & 0 \\ 0 & e^{16} & 0 & 0 \\ 0 & 0 & e^8 & 0 \\ 0 & 0 & 0 & e^{23} \end{pmatrix} \begin{pmatrix} -1 & -1 & 0 & 0 \\ 1 & -1 & 0 & 0 \\ 0 & 0 & -1 & -1 \\ 0 & 0 & 1 & -1 \end{pmatrix} \begin{pmatrix} 1 \\ 16 \\ 8 \\ 25 \end{pmatrix}$$
$$= \frac{1}{2}\begin{pmatrix} 17e - 15e^{16} \\ 17e + 15e^{16} \\ 33e^8 - 17e^{25} \\ 33e^8 + 17e^{25} \end{pmatrix}$$

To decrypt the cipher-block, once the system has received the cipher-blocks, it can decrypt by the reverse process.

To decrypt, the receiver receive $\mathbf{k}_{pub}$. Because he has $\mathbf{k}_{priv}$, he can compute $\mathbf{C} = \mathbf{k}_{priv} * \mathbf{k}_{pub}$.

To decrypt the cipher-block, once the system has received the cipher-blocks, it can decrypt by the reverse process:

$$T_1 = CY_1 = \frac{1}{2}\begin{pmatrix} 1 & -1 & 0 & 0 \\ 1 & 1 & 0 & 0 \\ 0 & 0 & 1 & -1 \\ 0 & 0 & 1 & 1 \end{pmatrix}\begin{pmatrix} 21e^{18} - 15e^3 \\ 21e^{18} + 15e^3 \\ 41e^{16} + 9e^{25} \\ 41e^{16} - 9e^{25} \end{pmatrix}$$

$$= \begin{pmatrix} -15e^3 \\ 21e^{18} \\ 9e^{25} \\ 41e^{16} \end{pmatrix}$$

$$T_2 = C^2Y_2 = \begin{pmatrix} 0 & -1 & 0 & 0 \\ 1 & 0 & 0 & 0 \\ 0 & 0 & 0 & -1 \\ 0 & 0 & 1 & 0 \end{pmatrix}\begin{pmatrix} 20e^{15} \\ 15e^{20} \\ 7e^{18} \\ 18e^7 \end{pmatrix} = \begin{pmatrix} 15e^{20} \\ 20e^{15} \\ -18e^7 \\ 7e^{18} \end{pmatrix}$$

$$T_3 = C^3Y_3 = \frac{1}{2}\begin{pmatrix} -1 & -1 & 0 & 0 \\ 1 & -1 & 0 & 0 \\ 0 & 0 & -1 & -1 \\ 0 & 0 & 1 & -1 \end{pmatrix}\begin{pmatrix} 17e - 15e^{16} \\ 17e + 15e^{16} \\ 33e^8 - 17e^{25} \\ 33e^8 + 17e^{25} \end{pmatrix}$$

$$= \begin{pmatrix} -17e \\ -15e^{16} \\ -33e^8 \\ -17e^{25} \end{pmatrix}$$

Because, to encrypt it was used the matrix **C** and its reverse, whereas to decrypt, only one time is necessary, some values can be negative. So, because all values will be considered positive, absolute values of components will be considered:

$$T_1 = CY_1 = \begin{pmatrix} 15e^3 \\ 21e^{18} \\ 9e^{25} \\ 41e^{16} \end{pmatrix}$$

$$T_2 = C^2Y_2 = \begin{pmatrix} 15e^{20} \\ 20e^{15} \\ 18e^7 \\ 7e^{18} \end{pmatrix}$$

$$T_3 = C^3Y_3 = \begin{pmatrix} 17e \\ 15e^{16} \\ 33e^8 \\ 17e^{25} \end{pmatrix}$$

Using the values of the table (see Appendix), for each value it can be found the exponent of exponential, so find the value of the plaintext. So, for $T_1$:

$15e^3$ : Exponent 3, so, the first character is "C"

$22e^{18}$: Exponent 18, so, "R"

$9e^{25}$: Exponent 25, so "Y"

$41e^{16}$: Exponent 16, so "P"

And so on.

Because each letter matched with a number, in the way that:

| A | B | C | D | E | F | G | H | I | J | K | L | M |
|---|---|---|---|---|---|---|---|---|---|---|---|---|
| 1 | 2 | 3 | 4 | 5 | 6 | 7 | 8 | 9 | 10 | 11 | 12 | 13 |
| N | O | P | Q | R | S | T | U | V | W | X | Y | Z |
| 14 | 15 | 16 | 17 | 18 | 19 | 20 | 21 | 22 | 23 | 24 | 25 | 26 |
| a | b | c | d | e | f | g | h | i | j | k | l | m |
| 27 | 28 | 29 | 30 | 31 | 32 | 33 | 34 | 35 | 36 | 37 | 38 | 39 |
| n | o | p | q | r | s | t | u | v | w | x | y | z |
| 40 | 41 | 42 | 43 | 44 | 45 | 46 | 47 | 48 | 49 | 50 | 51 | 52 |
| space | 0 | 1 | 2 | 3 | 4 | 5 | 6 | 7 | 8 | 9 | | |
| 53 | 54 | 55 | 56 | 57 | 58 | 59 | 60 | 61 | 62 | 63 | | |

The cipher-blocks will be:

$X_1 ==$ "CRYP"

$X_2 ==$ "TOGR"

$X_3 ==$ "APHY"

And the text "CRYPTGORAPHY" will be decrypted.

It could be possible to use a permutation matrix in the way to change the values of the output, and make more difficult an attack. So, for example, using the matrix P

$$P = \begin{pmatrix} 0 & 1 & 0 & 0 \\ 0 & 0 & 1 & 0 \\ 0 & 0 & 0 & 1 \\ 1 & 0 & 0 & 0 \end{pmatrix}$$

The output of the message encrypt, once permuted, will be:

Block 1: "CRYP"

$$Y_1^P = PC^{-1}DCX_1$$

$$= \frac{1}{2}\begin{pmatrix} 0 & 1 & 0 & 0 \\ 0 & 0 & 1 & 0 \\ 0 & 0 & 0 & 1 \\ 1 & 0 & 0 & 0 \end{pmatrix}\begin{pmatrix} 1 & 1 & 0 & 0 \\ -1 & 1 & 0 & 0 \\ 0 & 0 & 1 & 1 \\ 0 & 0 & -1 & 1 \end{pmatrix} *$$

$$* \begin{pmatrix} e^3 & 0 & 0 & 0 \\ 0 & e^{18} & 0 & 0 \\ 0 & 0 & e^{25} & 0 \\ 0 & 0 & 0 & e^{16} \end{pmatrix}\begin{pmatrix} 1 & -1 & 0 & 0 \\ 1 & 1 & 0 & 0 \\ 0 & 0 & 1 & -1 \\ 0 & 0 & 1 & 1 \end{pmatrix}\begin{pmatrix} 3 \\ 18 \\ 25 \\ 16 \end{pmatrix}$$

$$= \frac{1}{2}\begin{pmatrix} 21e^{18} + 15e^3 \\ 41e^{16} + 9e^{25} \\ 41e^{16} - 9e^{25} \\ 21e^{18} - 15e^3 \end{pmatrix}$$

# Orthogonal Matrix in Cryptography

Block 2: "TOGR"

$$Y_2^P = P^2(C^2)^{-1}DC^2X_2 = \begin{pmatrix} 0 & 0 & 1 & 0 \\ 0 & 0 & 0 & 1 \\ 1 & 0 & 0 & 0 \\ 0 & 1 & 0 & 0 \end{pmatrix} \begin{pmatrix} 0 & 1 & 0 & 0 \\ -1 & 0 & 0 & 0 \\ 0 & 0 & 0 & 1 \\ 0 & 0 & -1 & 0 \end{pmatrix} *$$

$$* \begin{pmatrix} e^{20} & 0 & 0 & 0 \\ 0 & e^{15} & 0 & 0 \\ 0 & 0 & e^7 & 0 \\ 0 & 0 & 0 & e^{18} \end{pmatrix} \begin{pmatrix} 0 & -1 & 0 & 0 \\ 1 & 0 & 0 & 0 \\ 0 & 0 & 0 & -1 \\ 0 & 0 & 1 & 0 \end{pmatrix} \begin{pmatrix} 20 \\ 15 \\ 7 \\ 18 \end{pmatrix} = \begin{pmatrix} 7e^{18} \\ 18e^7 \\ 20e^{15} \\ 15e^{20} \end{pmatrix}$$

Block 3: "APHY"

$$Y_3^P = P^3(C^3)^{-1}DC^3X_3$$

$$= \frac{1}{2} \begin{pmatrix} 0 & 0 & 0 & 1 \\ 1 & 0 & 0 & 0 \\ 0 & 1 & 0 & 0 \\ 0 & 0 & 1 & 0 \end{pmatrix} \begin{pmatrix} -1 & 1 & 0 & 0 \\ -1 & -1 & 0 & 0 \\ 0 & 0 & -1 & 1 \\ 0 & 0 & -1 & -1 \end{pmatrix}$$

$$* \begin{pmatrix} e & 0 & 0 & 0 \\ 0 & e^{16} & 0 & 0 \\ 0 & 0 & e^8 & 0 \\ 0 & 0 & 0 & e^{23} \end{pmatrix} \begin{pmatrix} -1 & -1 & 0 & 0 \\ 1 & -1 & 0 & 0 \\ 0 & 0 & -1 & -1 \\ 0 & 0 & 1 & -1 \end{pmatrix} \begin{pmatrix} 1 \\ 16 \\ 8 \\ 25 \end{pmatrix}$$

$$= \frac{1}{2} \begin{pmatrix} 33e^8 + 17e^{25} \\ 17e - 15e^{16} \\ 17e + 15e^{16} \\ 33e^8 - 17e^{25} \end{pmatrix}$$

To decrypt the message, let's take the inverse matrix:

$$P^{-1} = \begin{pmatrix} 0 & 0 & 0 & 1 \\ 1 & 0 & 0 & 0 \\ 0 & 1 & 0 & 0 \\ 0 & 0 & 1 & 0 \end{pmatrix}$$

$$P^{-2} = \begin{pmatrix} 0 & 0 & 1 & 0 \\ 0 & 0 & 0 & 1 \\ 1 & 0 & 0 & 0 \\ 0 & 1 & 0 & 0 \end{pmatrix}$$

$$P^{-3} = \begin{pmatrix} 0 & 1 & 0 & 0 \\ 0 & 0 & 1 & 0 \\ 0 & 0 & 0 & 1 \\ 1 & 0 & 0 & 0 \end{pmatrix}$$

$$Y_1 = P^{-1}C^{-1}DCX_1 = \frac{1}{2} \begin{pmatrix} 21e^{18} - 15e^3 \\ 21e^{18} + 15e^3 \\ 41e^{16} + 9e^{25} \\ 41e^{16} - 9e^{25} \end{pmatrix}$$

$$Y_2 = P^{-2}(C^2)^{-1}DC^2X_2 = \begin{pmatrix} 20e^{15} \\ 15e^{20} \\ 7e^{18} \\ 18e^7 \end{pmatrix}$$

$$Y_1 = P^{-3}(C^3)^{-1}DC^3X_3 = \frac{1}{2} \begin{pmatrix} 17e - 15e^{16} \\ 17e + 15e^{16} \\ 33e^8 - 17e^{25} \\ 33e^8 + 17e^{25} \end{pmatrix}$$

And, using the same procedure showed before, the system can find the vectors $T_i$.

So, the plaintext message will be:

"CRYPTOGRAPHY"

**ATTACK**

Suppose that attacker receives the encrypt text of "CRTYP"

$$Y_1 = PC^{-1}DCX_1 = \frac{1}{2}\begin{pmatrix} 21e^{18} + 15e^3 \\ 41e^{16} + 9e^{25} \\ 41e^{16} - 9e^{25} \\ 21e^{18} - 15e^3 \end{pmatrix}$$

Taking linear combinations:

$Y_{11}+Y_{14}=21e^{18}=1378859352$

$Y_{12}+Y_{23}=41e^{16}=364330531,3$

$Y_{12}-Y_{14}=15e^{3}=301,2830538$

$Y_{12}-Y_{23}=9e^{25}=648044094036,47$

Because the encryption is based on combinations, attacker must found a combination of exponentials to find 21 and 18 (and only if he knows that there's exponentials). Even so, the way would be trying several values of a and b in sense that, knowing c, $a*e^b=c$. But, the second block cipher is like:

$$Y_2 = P^2(C^2)^{-1}DC^2X_2$$

And the method used for decrypt the first block won't be possible in this case because in this case the matrix is $C^2$, so the attacker must use the same method but for a different matrix.

Because each character is codified by an exponential $e^{[char]}$, in sense that, A is codified by e, B, by $e^2$ and so on, there's not linear combinations, and that difficult much more an attack by frequencies. Also, the final cyphertext is a combination of several ciphers, so it will be very difficult to separate them (ex, $21e^{18} + 15e^3$ is the combination of encryption of letters "C" and "R"). The matrix permutation **P** changes the position of each cyphertext, and that makes more difficult to decrypt the message without **P**.

## EXTENSIONS

The method proposed can be extended to other function rather than exponential (hyperbolic sinus and hyperbolic cosines). Because there's a relationship between exponential function and hyperbolic functions, the same method could be applied.

It has shown that, for the exponential function,, the matrix is:

$$e^{\mathbf{A}} = \mathbf{C}^{-1} e^{D} \mathbf{C} = \mathbf{C}^{-1} \begin{pmatrix} e^{\lambda_1} & & & \\ & e^{\lambda_2} & & \\ & & \ddots & \\ & & & e^{\lambda_n} \end{pmatrix} \mathbf{C}$$

So, using hyperbolic sinus:

$$sinh\mathbf{A} = \frac{e^{\mathbf{A}} - e^{-\mathbf{A}}}{2} =$$

$$\mathbf{C}^{-1} \begin{pmatrix} \frac{e^{\lambda_1} - e^{-\lambda_1}}{2} & & & \\ & \frac{e^{\lambda_2} - e^{-\lambda_2}}{2} & & \\ & & \ddots & \\ & & & \frac{e^{\lambda_n} - e^{-\lambda_n}}{2} \end{pmatrix} \mathbf{C}$$

$$sinh\mathbf{A} = \mathbf{C}^{-1} \begin{pmatrix} sinh\lambda_1 & & & \\ & sinh\lambda_2 & & \\ & & \ddots & \\ & & & sinh\lambda_n \end{pmatrix} \mathbf{C}$$

Because the sinh is not periodic in $\Re$, it can be also used as the same method seen before.

## CONCLUSSIONS

In this work, it has showed how to use Hermitian matrix to encrypt and decrypt messages by cipher-block. Because each cipher-block is a linear combination of several cipher-blocks, it makes much difficult to plain an attack. Using the properties of hermitian matrix, it will be faster to calculate the reverse of the matrix to generate the ciphertext and the plaintext.

Orthogonal Matrix in Cryptography

Values of file*e$^{col}$

| f*e$^c$ | 1 | 2 | 3 | 4 | 5 | 6 | 7 | 8 | 9 | 10 | 11 | 12 | 13 | 14 | 15 |
|---|---|---|---|---|---|---|---|---|---|---|---|---|---|---|---|
| 1 | 2,72 | 7,39 | 20,09 | 54,60 | 148,41 | 403,43 | 1096,63 | 2980,96 | 8103,08 | 22026,47 | 59874,14 | 162754,79 | 442413,39 | 1202604,28 | 3269017,37 |
| 2 | 5,44 | 14,78 | 40,17 | 109,20 | 296,83 | 806,86 | 2193,27 | 5961,92 | 16206,17 | 44052,93 | 119748,28 | 325509,58 | 884826,78 | 2405208,57 | 6538034,74 |
| 3 | 8,15 | 22,17 | 60,26 | 163,79 | 445,24 | 1210,29 | 3289,90 | 8942,87 | 24309,25 | 66079,40 | 179622,43 | 488264,37 | 1327240,18 | 3607812,85 | 9807052,12 |
| 4 | 10,87 | 29,56 | 80,34 | 218,39 | 593,65 | 1613,72 | 4386,53 | 11923,83 | 32412,34 | 88105,86 | 239496,57 | 651019,17 | 1769653,57 | 4810417,14 | 13076069,49 |
| 5 | 13,59 | 36,95 | 100,43 | 272,99 | 742,07 | 2017,14 | 5483,17 | 14904,79 | 40515,42 | 110132,33 | 299370,71 | 813773,96 | 2212066,96 | 6013021,42 | 16345086,86 |
| 6 | 16,31 | 44,33 | 120,51 | 327,59 | 890,48 | 2420,57 | 6579,80 | 17885,75 | 48618,50 | 132158,79 | 359244,85 | 976528,75 | 2654480,35 | 7215625,70 | 19614104,23 |
| 7 | 19,03 | 51,72 | 140,60 | 382,19 | 1038,89 | 2824,00 | 7676,43 | 20866,71 | 56721,59 | 154185,26 | 419118,99 | 1139283,54 | 3096893,74 | 8418229,99 | 22883121,61 |
| 8 | 21,75 | 59,11 | 160,68 | 436,79 | 1187,31 | 3227,43 | 8773,07 | 23847,66 | 64824,67 | 176211,73 | 478993,13 | 1302038,33 | 3539307,14 | 9620834,27 | 26152138,98 |
| 9 | 24,46 | 66,50 | 180,77 | 491,38 | 1335,72 | 3630,86 | 9869,70 | 26828,62 | 72927,76 | 198238,19 | 538867,28 | 1464793,12 | 3981720,53 | 10823438,56 | 29421156,35 |
| 10 | 27,18 | 73,89 | 200,86 | 545,98 | 1484,13 | 4034,29 | 10966,33 | 29809,58 | 81030,84 | 220264,66 | 598741,42 | 1627547,91 | 4424133,92 | 12026042,84 | 32690173,72 |
| 11 | 29,90 | 81,28 | 220,94 | 600,58 | 1632,54 | 4437,72 | 12062,96 | 32790,54 | 89133,92 | 242291,12 | 658615,56 | 1790302,71 | 4866547,31 | 13228647,13 | 35959191,10 |
| 12 | 32,62 | 88,67 | 241,03 | 655,18 | 1780,96 | 4841,15 | 13159,60 | 35771,50 | 97237,01 | 264317,59 | 718489,70 | 1953057,50 | 5308960,70 | 14431251,41 | 39228208,47 |
| 13 | 35,34 | 96,06 | 261,11 | 709,78 | 1929,37 | 5244,57 | 14256,23 | 38752,45 | 105340,09 | 286344,06 | 778363,84 | 2115812,29 | 5751374,10 | 15633855,69 | 42497225,84 |
| 14 | 38,06 | 103,45 | 281,20 | 764,37 | 2077,78 | 5648,00 | 15352,86 | 41733,41 | 113443,17 | 308370,52 | 838237,98 | 2278567,08 | 6193787,49 | 16836459,98 | 45766243,21 |
| 15 | 40,77 | 110,84 | 301,28 | 818,97 | 2226,20 | 6051,43 | 16449,50 | 44714,37 | 121546,26 | 330396,99 | 898112,13 | 2441321,87 | 6636200,88 | 18039064,26 | 49035260,59 |
| 16 | 43,49 | 118,22 | 321,37 | 873,57 | 2374,61 | 6454,86 | 17546,13 | 47695,33 | 129649,34 | 352423,45 | 957986,27 | 2604076,66 | 7078614,27 | 19241668,55 | 52304277,96 |
| 17 | 46,21 | 125,61 | 341,45 | 928,17 | 2523,02 | 6858,29 | 18642,76 | 50676,29 | 137752,43 | 374449,92 | 1017860,41 | 2766831,45 | 7521027,66 | 20444272,83 | 55573295,33 |
| 18 | 48,93 | 133,00 | 361,54 | 982,77 | 2671,44 | 7261,72 | 19739,40 | 53657,24 | 145855,51 | 396476,38 | 1077734,55 | 2929586,25 | 7963441,06 | 21646877,11 | 58842312,70 |
| 19 | 51,65 | 140,39 | 381,63 | 1037,36 | 2819,85 | 7665,15 | 20836,03 | 56638,20 | 153958,59 | 418502,85 | 1137608,69 | 3092341,04 | 8405854,45 | 22849481,40 | 62111330,08 |
| 20 | 54,37 | 147,78 | 401,71 | 1091,96 | 2968,26 | 8068,58 | 21932,66 | 59619,16 | 162061,68 | 440529,32 | 1197482,83 | 3255095,83 | 8848267,84 | 24052085,68 | 65380347,45 |
| 21 | 57,08 | 155,17 | 421,80 | 1146,56 | 3116,68 | 8472,00 | 23029,30 | 62600,12 | 170164,76 | 462555,78 | 1257356,98 | 3417850,62 | 9290681,23 | 25254689,97 | 68649364,82 |
| 22 | 59,80 | 162,56 | 441,88 | 1201,16 | 3265,09 | 8875,43 | 24125,93 | 65581,08 | 178267,85 | 484582,25 | 1317231,12 | 3580605,41 | 9733094,62 | 26457294,25 | 71918382,19 |
| 23 | 62,52 | 169,95 | 461,97 | 1255,76 | 3413,50 | 9278,86 | 25222,56 | 68562,03 | 186370,93 | 506608,71 | 1377105,26 | 3743360,20 | 10175508,02 | 27659898,54 | 75187399,57 |
| 24 | 65,24 | 177,34 | 482,05 | 1310,36 | 3561,92 | 9682,29 | 26319,20 | 71542,99 | 194474,01 | 528635,18 | 1436979,40 | 3906114,99 | 10617921,41 | 28862502,82 | 78456416,94 |
| 25 | 67,96 | 184,73 | 502,14 | 1364,95 | 3710,33 | 10085,72 | 27415,83 | 74523,95 | 202577,10 | 550661,64 | 1496853,54 | 4068869,79 | 11060334,80 | 30065107,10 | 81725434,31 |
| 26 | 70,68 | 192,12 | 522,22 | 1419,55 | 3858,74 | 10489,15 | 28512,46 | 77504,91 | 210680,18 | 572688,11 | 1556727,68 | 4231624,58 | 11502748,19 | 31267711,39 | 84994451,68 |
| 27 | 73,39 | 199,50 | 542,31 | 1474,15 | 4007,16 | 10892,58 | 29609,10 | 80485,87 | 218783,27 | 594714,58 | 1616601,83 | 4394379,37 | 11945161,58 | 32470315,67 | 88263469,06 |
| 28 | 76,11 | 206,89 | 562,40 | 1528,75 | 4155,57 | 11296,01 | 30705,73 | 83466,82 | 226886,35 | 616741,04 | 1676475,97 | 4557134,16 | 12387574,98 | 33672919,96 | 91532486,43 |
| 29 | 78,83 | 214,28 | 582,48 | 1583,35 | 4303,98 | 11699,44 | 31802,36 | 86447,78 | 234989,43 | 638767,51 | 1736350,11 | 4719888,95 | 12829988,37 | 34875524,24 | 94801503,80 |

Values of file*e$^{col}$

| f*e$^c$ | 1 | 2 | 3 | 4 | 5 | 6 | 7 | 8 | 9 | 10 | 11 | 12 | 13 | 14 | 15 |
|---|---|---|---|---|---|---|---|---|---|---|---|---|---|---|---|
| 30 | 81,55 | 221,67 | 602,57 | 1637,94 | 4452,39 | 12102,86 | 32898,99 | 89428,74 | 243092,52 | 660793,97 | 1796224,25 | 4882643,74 | 13272401,76 | 36078128,52 | 98070521,17 |
| 31 | 84,27 | 229,06 | 622,65 | 1692,54 | 4600,81 | 12506,29 | 33995,63 | 92409,70 | 251195,60 | 682820,44 | 1856098,39 | 5045398,53 | 13714815,15 | 37280732,81 | 101339538,55 |
| 32 | 86,99 | 236,45 | 642,74 | 1747,14 | 4749,22 | 12909,72 | 35092,26 | 95390,66 | 259298,69 | 704846,91 | 1915972,53 | 5208153,33 | 14157228,54 | 38483337,09 | 104608555,92 |
| 33 | 89,70 | 243,84 | 662,82 | 1801,74 | 4897,63 | 13313,15 | 36188,89 | 98371,61 | 267401,77 | 726873,37 | 1975846,68 | 5370908,12 | 14599641,94 | 39685941,38 | 107877573,29 |
| 34 | 92,42 | 251,23 | 682,91 | 1856,34 | 5046,05 | 13716,58 | 37285,53 | 101352,57 | 275504,85 | 748899,84 | 2035720,82 | 5533662,91 | 15042055,33 | 40888545,66 | 111146590,66 |
| 35 | 95,14 | 258,62 | 702,99 | 1910,94 | 5194,46 | 14120,01 | 38382,16 | 104333,53 | 283607,94 | 770926,30 | 2095594,96 | 5696417,70 | 15484468,72 | 42091149,95 | 114415608,04 |
| 36 | 97,86 | 266,01 | 723,08 | 1965,53 | 5342,87 | 14523,44 | 39478,79 | 107314,49 | 291711,02 | 792952,77 | 2155469,10 | 5859172,49 | 15926882,11 | 43293754,23 | 117684625,41 |
| 37 | 100,58 | 273,40 | 743,16 | 2020,13 | 5491,29 | 14926,87 | 40575,43 | 110295,45 | 299814,11 | 814979,23 | 2215343,24 | 6021927,28 | 16369295,50 | 44496358,51 | 120953642,78 |
| 38 | 103,29 | 280,78 | 763,25 | 2074,73 | 5639,70 | 15330,29 | 41672,06 | 113276,40 | 307917,19 | 837005,70 | 2275217,39 | 6184682,07 | 16811708,90 | 45698962,80 | 124222660,15 |
| 39 | 106,01 | 288,17 | 783,34 | 2129,33 | 5788,11 | 15733,72 | 42768,69 | 116257,36 | 316020,27 | 859032,17 | 2335091,53 | 6347436,87 | 17254122,29 | 46901567,08 | 127491677,53 |
| 40 | 108,73 | 295,56 | 803,42 | 2183,93 | 5936,53 | 16137,15 | 43865,33 | 119238,32 | 324123,36 | 881058,63 | 2394965,67 | 6510191,66 | 17696535,68 | 48104171,37 | 130760694,90 |
| 41 | 111,45 | 302,95 | 823,51 | 2238,52 | 6084,94 | 16540,58 | 44961,96 | 122219,28 | 332226,44 | 903085,10 | 2454839,81 | 6672946,45 | 18138949,07 | 49306775,65 | 134029712,27 |
| 42 | 114,17 | 310,34 | 843,59 | 2293,12 | 6233,35 | 16944,01 | 46058,59 | 125200,24 | 340329,52 | 925111,56 | 2514713,95 | 6835701,24 | 18581362,46 | 50509379,93 | 137298729,64 |
| 43 | 116,89 | 317,73 | 863,68 | 2347,72 | 6381,77 | 17347,44 | 47155,23 | 128181,19 | 348432,61 | 947138,03 | 2574588,09 | 6998456,03 | 19023775,86 | 51711984,22 | 140567747,02 |
| 44 | 119,60 | 325,12 | 883,76 | 2402,32 | 6530,18 | 17750,87 | 48251,86 | 131162,15 | 356535,69 | 969164,49 | 2634462,24 | 7161210,82 | 19466189,25 | 52914588,50 | 143836764,39 |
| 45 | 122,32 | 332,51 | 903,85 | 2456,92 | 6678,59 | 18154,30 | 49348,49 | 134143,11 | 364638,78 | 991190,96 | 2694336,38 | 7323965,61 | 19908602,64 | 54117192,79 | 147105781,76 |
| 46 | 125,04 | 339,90 | 923,93 | 2511,51 | 6827,01 | 18557,72 | 50445,13 | 137124,07 | 372741,86 | 1013217,43 | 2754210,52 | 7486720,41 | 20351016,03 | 55319797,07 | 150374799,13 |
| 47 | 127,76 | 347,29 | 944,02 | 2566,11 | 6975,42 | 18961,15 | 51541,76 | 140105,03 | 380844,94 | 1035243,89 | 2814084,66 | 7649475,20 | 20793429,42 | 56522401,36 | 153643816,51 |
| 48 | 130,48 | 354,67 | 964,11 | 2620,71 | 7123,83 | 19364,58 | 52638,39 | 143085,98 | 388948,03 | 1057270,36 | 2873958,80 | 7812229,99 | 21235842,82 | 57725005,64 | 156912833,88 |
| 49 | 133,20 | 362,06 | 984,19 | 2675,31 | 7272,24 | 19768,01 | 53735,02 | 146066,94 | 397051,11 | 1079296,82 | 2933832,94 | 7974984,78 | 21678256,21 | 58927609,92 | 160181851,25 |
| 50 | 135,91 | 369,45 | 1004,28 | 2729,91 | 7420,66 | 20171,44 | 54831,66 | 149047,90 | 405154,20 | 1101323,29 | 2993707,09 | 8137739,57 | 22120669,60 | 60130214,21 | 163450868,62 |
| 51 | 138,63 | 376,84 | 1024,36 | 2784,51 | 7569,07 | 20574,87 | 55928,29 | 152028,86 | 413257,28 | 1123349,76 | 3053581,23 | 8300494,36 | 22563082,99 | 61332818,49 | 166719886,00 |
| 52 | 141,35 | 384,23 | 1044,45 | 2839,10 | 7717,48 | 20978,30 | 57024,92 | 155009,82 | 421360,36 | 1145376,22 | 3113455,37 | 8463249,15 | 23005496,38 | 62535422,78 | 169988903,37 |
| 53 | 144,07 | 391,62 | 1064,53 | 2893,70 | 7865,90 | 21381,73 | 58121,56 | 157990,77 | 429463,45 | 1167402,69 | 3173329,51 | 8626003,95 | 23447909,78 | 63738027,06 | 173257920,74 |
| 54 | 146,79 | 399,01 | 1084,62 | 2948,30 | 8014,31 | 21785,15 | 59218,19 | 160971,73 | 437566,53 | 1189429,15 | 3233203,65 | 8788758,74 | 23890323,17 | 64940631,34 | 176526938,11 |
| 55 | 149,51 | 406,40 | 1104,70 | 3002,90 | 8162,72 | 22188,58 | 60314,82 | 163952,69 | 445669,62 | 1211455,62 | 3293077,79 | 8951513,53 | 24332736,56 | 66143235,63 | 179795955,49 |
| 56 | 152,22 | 413,79 | 1124,79 | 3057,50 | 8311,14 | 22592,01 | 61411,46 | 166933,65 | 453772,70 | 1233482,08 | 3352951,94 | 9114268,32 | 24775149,95 | 67345839,91 | 183064972,86 |
| 57 | 154,94 | 421,18 | 1144,88 | 3112,09 | 8459,55 | 22995,44 | 62508,09 | 169914,61 | 461875,78 | 1255508,55 | 3412826,08 | 9277023,11 | 25217563,34 | 68548444,20 | 186333990,23 |
| 58 | 157,66 | 428,57 | 1164,96 | 3166,69 | 8607,96 | 23398,87 | 63604,72 | 172895,56 | 469978,87 | 1277535,02 | 3472700,22 | 9439777,90 | 25659976,74 | 69751048,48 | 189603007,60 |

Values of file*e$^{col}$

| | | | | | | | | | | | | | | | |
|---|---|---|---|---|---|---|---|---|---|---|---|---|---|---|---|
| 59 | 160,38 | 435,95 | 1185,05 | 3221,29 | 8756,38 | 23802,30 | 64701,36 | 175876,52 | 478081,95 | 1299561,48 | 3532574,36 | 9602532,69 | 26102390,13 | 70953652,77 | 192872024,98 |
| 60 | 163,10 | 443,34 | 1205,13 | 3275,89 | 8904,79 | 24205,73 | 65797,99 | 178857,48 | 486185,04 | 1321587,95 | 3592448,50 | 9765287,49 | 26544803,52 | 72156257,05 | 196141042,35 |
| 61 | 165,82 | 450,73 | 1225,22 | 3330,49 | 9053,20 | 24609,16 | 66894,62 | 181838,44 | 494288,12 | 1343614,41 | 3652322,64 | 9928042,28 | 26987216,91 | 73358861,33 | 199410059,72 |
| 62 | 168,53 | 458,12 | 1245,30 | 3385,09 | 9201,62 | 25012,59 | 67991,26 | 184819,40 | 502391,20 | 1365640,88 | 3712196,79 | 10090797,07 | 27429630,30 | 74561465,62 | 202679077,09 |
| 63 | 171,25 | 465,51 | 1265,39 | 3439,68 | 9350,03 | 25416,01 | 69087,89 | 187800,35 | 510494,29 | 1387667,35 | 3772070,93 | 10253551,86 | 27872043,70 | 75764069,90 | 205948094,47 |
| 64 | 173,97 | 472,90 | 1285,47 | 3494,28 | 9498,44 | 25819,44 | 70184,52 | 190781,31 | 518597,37 | 1409693,81 | 3831945,07 | 10416306,65 | 28314457,09 | 76966674,19 | 209217111,84 |

| f*e$^c$ | 16 | 17 | 18 | 19 | 20 | 21 | 22 | 23 | 24 | 25 | 26 |
|---|---|---|---|---|---|---|---|---|---|---|---|
| 1 | 8886110,52 | 24154952,8 | 65659969,1 | 178482301 | 485165195 | 1318815734 | 3584912846 | 9744803446 | 26489122130 | 72004899337 | 1,9573E+11 |
| 2 | 17772221 | 48309905,5 | 131319938 | 356964602 | 970330391 | 2637631469 | 7169825692 | 19489606892 | 52978244260 | 1,4401E+11 | 3,91459E+11 |
| 3 | 26658331,6 | 72464858,3 | 196979907 | 535446903 | 1455495586 | 3956447203 | 10754738538 | 29234410339 | 79467366390 | 2,16015E+11 | 5,87189E+11 |
| 4 | 35544442,1 | 96619811 | 262639877 | 713929204 | 1940660782 | 5275262938 | 14339651385 | 38979213785 | 1,05956E+11 | 2,8802E+11 | 7,82918E+11 |
| 5 | 44430552,6 | 120774764 | 328299846 | 892411505 | 2425825977 | 6594078672 | 17924564231 | 48724017231 | 1,32446E+11 | 3,60024E+11 | 9,78648E+11 |
| 6 | 53316663,1 | 144929717 | 393959815 | 1070893806 | 2910991172 | 7912894407 | 21509477077 | 58468820677 | 1,58935E+11 | 4,32029E+11 | 1,17438E+12 |
| 7 | 62202773,6 | 169084669 | 459619784 | 1249376107 | 3396156368 | 9231710141 | 25094389923 | 68213624124 | 1,85424E+11 | 5,04034E+11 | 1,37011E+12 |
| 8 | 71088884,2 | 193239622 | 525279753 | 1427858408 | 3881321563 | 10550525876 | 28679302769 | 77958427570 | 2,11913E+11 | 5,76039E+11 | 1,56584E+12 |
| 9 | 79974994,7 | 217394575 | 590939722 | 1606340709 | 4366486759 | 11869341610 | 32264215615 | 87703231016 | 2,38402E+11 | 6,48044E+11 | 1,76157E+12 |
| 10 | 88861105,2 | 241549528 | 656599691 | 1784823010 | 4851651954 | 13188157345 | 35849128461 | 97448034462 | 2,64891E+11 | 7,20049E+11 | 1,9573E+12 |
| 11 | 97747215,7 | 265704480 | 722259661 | 1963305311 | 5336817150 | 14506973079 | 39434041307 | 1,07193E+11 | 2,9138E+11 | 7,92054E+11 | 2,15303E+12 |
| 12 | 106633326 | 289859433 | 787919630 | 2141787612 | 5821982345 | 15825788814 | 43018954154 | 1,16938E+11 | 3,17869E+11 | 8,64059E+11 | 2,34876E+12 |
| 13 | 115519437 | 314014386 | 853579599 | 2320269913 | 6307147540 | 17144604548 | 46603867000 | 1,26682E+11 | 3,44359E+11 | 9,36064E+11 | 2,54448E+12 |
| 14 | 124405547 | 338169339 | 919239568 | 2498752213 | 6792312736 | 18463420283 | 50188779846 | 1,36427E+11 | 3,70848E+11 | 1,00807E+12 | 2,74021E+12 |
| 15 | 133291658 | 362324291 | 984899537 | 2677234514 | 7277477931 | 19782236017 | 53773692692 | 1,46172E+11 | 3,97337E+11 | 1,08007E+12 | 2,93594E+12 |
| 16 | 142177768 | 386479244 | 1050559506 | 2855716815 | 7762643127 | 21101051752 | 57358605538 | 1,55917E+11 | 4,23826E+11 | 1,15208E+12 | 3,13167E+12 |
| 17 | 151063879 | 410634197 | 1116219475 | 3034199116 | 8247808322 | 22419867486 | 60943518384 | 1,65662E+11 | 4,50315E+11 | 1,22408E+12 | 3,3274E+12 |
| 18 | 159949989 | 434789150 | 1181879444 | 3212681417 | 8732973517 | 23738683221 | 64528431230 | 1,75406E+11 | 4,76804E+11 | 1,29609E+12 | 3,52313E+12 |
| 19 | 168836100 | 458944102 | 1247539414 | 3391163718 | 9218138713 | 25057498955 | 68113344077 | 1,85151E+11 | 5,03293E+11 | 1,36809E+12 | 3,71886E+12 |
| 20 | 177722210 | 483099055 | 1313199383 | 3569646019 | 9703303908 | 26376314690 | 71698256923 | 1,94896E+11 | 5,29782E+11 | 1,4401E+12 | 3,91459E+12 |
| 21 | 186608321 | 507254008 | 1378859352 | 3748128320 | 1,0188E+10 | 27695130424 | 75283169769 | 2,04641E+11 | 5,56272E+11 | 1,5121E+12 | 4,11032E+12 |

Values of file*e^col

| f*e^c | 16 | 17 | 18 | 19 | 20 | 21 | 22 | 23 | 24 | 25 | 26 |
|---|---|---|---|---|---|---|---|---|---|---|---|
| 22 | 195494431 | 531408961 | 1444519321 | 3926610621 | 1,0674E+10 | 29013946159 | 78868082615 | 2,14386E+11 | 5,82761E+11 | 1,58411E+12 | 4,30605E+12 |
| 23 | 204380542 | 555563913 | 1510179290 | 4105092922 | 1,1159E+10 | 30332761893 | 82452995461 | 2,2413E+11 | 6,0925E+11 | 1,65611E+12 | 4,50178E+12 |
| 24 | 213266652 | 579718866 | 1575839259 | 4283575223 | 1,1644E+10 | 31651577628 | 86037908307 | 2,33875E+11 | 6,35739E+11 | 1,72812E+12 | 4,69751E+12 |
| 25 | 222152763 | 603873819 | 1641499228 | 4462057524 | 1,2129E+10 | 32970393362 | 89622821153 | 2,4362E+11 | 6,62228E+11 | 1,80012E+12 | 4,89324E+12 |
| 26 | 231038874 | 628028772 | 1707159198 | 4640539825 | 1,2614E+10 | 34289209097 | 93207733999 | 2,53365E+11 | 6,88717E+11 | 1,87213E+12 | 5,08897E+12 |
| 27 | 239924984 | 652183724 | 1772819167 | 4819022126 | 1,3099E+10 | 35608024831 | 96792646846 | 2,6311E+11 | 7,15206E+11 | 1,94413E+12 | 5,2847E+12 |
| 28 | 248811095 | 676338677 | 1838479136 | 4997504427 | 1,3585E+10 | 36926840566 | 1,00378E+11 | 2,72854E+11 | 7,41695E+11 | 2,01614E+12 | 5,48043E+12 |
| 29 | 257697205 | 700493630 | 1904139105 | 5175986728 | 1,407E+10 | 38245656300 | 1,03962E+11 | 2,82599E+11 | 7,68185E+11 | 2,08814E+12 | 5,67616E+12 |
| 30 | 266583316 | 724648583 | 1969799074 | 5354469029 | 1,4555E+10 | 39564472034 | 1,07547E+11 | 2,92344E+11 | 7,94674E+11 | 2,16015E+12 | 5,87189E+12 |
| 31 | 275469426 | 748803535 | 2035459043 | 5532951330 | 1,504E+10 | 40883287769 | 1,11132E+11 | 3,02089E+11 | 8,21163E+11 | 2,23215E+12 | 6,06762E+12 |
| 32 | 284355537 | 772958488 | 2101119012 | 5711433631 | 1,5525E+10 | 42202103503 | 1,14717E+11 | 3,11834E+11 | 8,47652E+11 | 2,30416E+12 | 6,26335E+12 |
| 33 | 293241647 | 797113441 | 2166778982 | 5889915932 | 1,601E+10 | 43520919238 | 1,18302E+11 | 3,21579E+11 | 8,74141E+11 | 2,37616E+12 | 6,45908E+12 |
| 34 | 302127758 | 821268394 | 2232438951 | 6068398233 | 1,6496E+10 | 44839734972 | 1,21887E+11 | 3,31323E+11 | 9,0063E+11 | 2,44817E+12 | 6,65481E+12 |
| 35 | 311013868 | 845423346 | 2298098920 | 6246880534 | 1,6981E+10 | 46158550707 | 1,25472E+11 | 3,41068E+11 | 9,27119E+11 | 2,52017E+12 | 6,85054E+12 |
| 36 | 319899979 | 869578299 | 2363758889 | 6425362835 | 1,7466E+10 | 47477366441 | 1,29057E+11 | 3,50813E+11 | 9,53608E+11 | 2,59218E+12 | 7,04627E+12 |
| 37 | 328786089 | 893733252 | 2429418858 | 6603845136 | 1,7951E+10 | 48796182176 | 1,32642E+11 | 3,60558E+11 | 9,80098E+11 | 2,66418E+12 | 7,242E+12 |
| 38 | 337672200 | 917888205 | 2495078827 | 6782327437 | 1,8436E+10 | 50114997910 | 1,36227E+11 | 3,70303E+11 | 1,00659E+12 | 2,73619E+12 | 7,43773E+12 |
| 39 | 346558310 | 942043157 | 2560738796 | 6960809738 | 1,8921E+10 | 51433813645 | 1,39812E+11 | 3,80047E+11 | 1,03308E+12 | 2,80819E+12 | 7,63345E+12 |
| 40 | 355444421 | 966198110 | 2626398765 | 7139292039 | 1,9407E+10 | 52752629379 | 1,43397E+11 | 3,89792E+11 | 1,05956E+12 | 2,8802E+12 | 7,82918E+12 |
| 41 | 364330531 | 990353063 | 2692058735 | 7317774339 | 1,9892E+10 | 54071445114 | 1,46981E+11 | 3,99537E+11 | 1,08605E+12 | 2,9522E+12 | 8,02491E+12 |
| 42 | 373216642 | 1014508016 | 2757718704 | 7496256640 | 2,0377E+10 | 55390260848 | 1,50566E+11 | 4,09282E+11 | 1,11254E+12 | 3,02421E+12 | 8,22064E+12 |
| 43 | 382102752 | 1038662968 | 2823378673 | 7674738941 | 2,0862E+10 | 56709076583 | 1,54151E+11 | 4,19027E+11 | 1,13903E+12 | 3,09621E+12 | 8,41637E+12 |
| 44 | 390988863 | 1062817921 | 2889038642 | 7853221242 | 2,1347E+10 | 58027892317 | 1,57736E+11 | 4,28771E+11 | 1,16552E+12 | 3,16822E+12 | 8,6121E+12 |
| 45 | 399874973 | 1086972874 | 2954698611 | 8031703543 | 2,1832E+10 | 59346708052 | 1,61321E+11 | 4,38516E+11 | 1,19201E+12 | 3,24022E+12 | 8,80783E+12 |
| 46 | 408761084 | 1111127827 | 3020358580 | 8210185844 | 2,2318E+10 | 60665523786 | 1,64906E+11 | 4,48261E+11 | 1,2185E+12 | 3,31223E+12 | 9,00356E+12 |
| 47 | 417647194 | 1135282779 | 3086018549 | 8388668145 | 2,2803E+10 | 61984339521 | 1,68491E+11 | 4,58006E+11 | 1,24499E+12 | 3,38423E+12 | 9,19929E+12 |
| 48 | 426533305 | 1159437732 | 3151678519 | 8567150446 | 2,3288E+10 | 63303155255 | 1,72076E+11 | 4,67751E+11 | 1,27148E+12 | 3,45624E+12 | 9,39502E+12 |
| 49 | 435419416 | 1183592685 | 3217338488 | 8745632747 | 2,3773E+10 | 64621970990 | 1,75661E+11 | 4,77495E+11 | 1,29797E+12 | 3,52824E+12 | 9,59075E+12 |
| 50 | 444305526 | 1207747638 | 3282998457 | 8924115048 | 2,4258E+10 | 65940786724 | 1,79246E+11 | 4,8724E+11 | 1,32446E+12 | 3,60024E+12 | 9,78648E+12 |

Values of file*e^(col)

| f*e^c | 16 | 17 | 18 | 19 | 20 | 21 | 22 | 23 | 24 | 25 | 26 |
|---|---|---|---|---|---|---|---|---|---|---|---|
| 51 | 453191637 | 1231902590 | 3348658426 | 9102597349 | 2,4743E+10 | 67259602459 | 1,82831E+11 | 4,96985E+11 | 1,35095E+12 | 3,67225E+12 | 9,98221E+12 |
| 52 | 462077747 | 1256057543 | 3414318395 | 9281079650 | 2,5229E+10 | 68578418193 | 1,86415E+11 | 5,0673E+11 | 1,37743E+12 | 3,74425E+12 | 1,01779E+13 |
| 53 | 470963858 | 1280212496 | 3479978364 | 9459561951 | 2,5714E+10 | 69897233928 | 1,9E+11 | 5,16475E+11 | 1,40392E+12 | 3,81626E+12 | 1,03737E+13 |
| 54 | 479849968 | 1304367449 | 3545638333 | 9638044252 | 2,6199E+10 | 71216049662 | 1,93585E+11 | 5,26219E+11 | 1,43041E+12 | 3,88826E+12 | 1,05694E+13 |
| 55 | 488736079 | 1328522401 | 3611298303 | 9816526553 | 2,6684E+10 | 72534865397 | 1,9717E+11 | 5,35964E+11 | 1,4569E+12 | 3,96027E+12 | 1,07651E+13 |
| 56 | 497622189 | 1352677354 | 3676958272 | 9995008854 | 2,7169E+10 | 73853681131 | 2,00755E+11 | 5,45709E+11 | 1,48339E+12 | 4,03227E+12 | 1,09609E+13 |
| 57 | 506508300 | 1376832307 | 3742618241 | 1,0173E+10 | 2,7654E+10 | 75172496866 | 2,0434E+11 | 5,55454E+11 | 1,50988E+12 | 4,10428E+12 | 1,11566E+13 |
| 58 | 515394410 | 1400987260 | 3808278210 | 1,0352E+10 | 2,814E+10 | 76491312600 | 2,07925E+11 | 5,65199E+11 | 1,53637E+12 | 4,17628E+12 | 1,13523E+13 |
| 59 | 524280521 | 1425142212 | 3873938179 | 1,053E+10 | 2,8625E+10 | 77810128335 | 2,1151E+11 | 5,74943E+11 | 1,56286E+12 | 4,24829E+12 | 1,1548E+13 |
| 60 | 533166631 | 1449297165 | 3939598148 | 1,0709E+10 | 2,911E+10 | 79128944069 | 2,15095E+11 | 5,84688E+11 | 1,58935E+12 | 4,32029E+12 | 1,17438E+13 |
| 61 | 542052742 | 1473452118 | 4005258117 | 1,0887E+10 | 2,9595E+10 | 80447759803 | 2,1868E+11 | 5,94433E+11 | 1,61584E+12 | 4,3923E+12 | 1,19395E+13 |
| 62 | 550938852 | 1497607071 | 4070918087 | 1,1066E+10 | 3,008E+10 | 81766575538 | 2,22265E+11 | 6,04178E+11 | 1,64233E+12 | 4,4643E+12 | 1,21352E+13 |
| 63 | 559824963 | 1521762023 | 4136578056 | 1,1244E+10 | 3,0565E+10 | 83085391272 | 2,2585E+11 | 6,13923E+11 | 1,66881E+12 | 4,53631E+12 | 1,2331E+13 |
| 64 | 568711073 | 1545916976 | 4202238025 | 1,1423E+10 | 3,1051E+10 | 84404207007 | 2,29434E+11 | 6,23667E+11 | 1,6953E+12 | 4,60831E+12 | 1,25267E+13 |

| f*e^c | 27 | 28 | 29 | 30 | 31 | 32 | 33 | 34 | 35 | 36 | 37 | 38 |
|---|---|---|---|---|---|---|---|---|---|---|---|---|
| 1 | 5,3205E+11 | 1,4463E+12 | 3,9313E+12 | 1,0686E+13 | 2,9049E+13 | 7,8963E+13 | 2,14644E+14 | 5,83462E+14 | 1,58601E+15 | 4,31123E+15 | 1,17191E+16 | 3,1856E+16 |
| 2 | 1,0641E+12 | 2,8925E+12 | 7,8627E+12 | 2,1373E+13 | 5,8098E+13 | 1,57926E+14 | 4,29287E+14 | 1,16692E+15 | 3,17203E+15 | 8,62246E+15 | 2,34383E+16 | 6,3712E+16 |
| 3 | 1,5961E+12 | 4,3388E+12 | 1,1794E+13 | 3,2059E+13 | 8,7147E+13 | 2,36889E+14 | 6,43931E+14 | 1,75039E+15 | 4,75804E+15 | 1,29337E+16 | 3,51574E+16 | 9,5568E+16 |
| 4 | 2,1282E+12 | 5,785E+12 | 1,5725E+13 | 4,2746E+13 | 1,162E+14 | 3,15852E+14 | 8,58574E+14 | 2,33385E+15 | 6,34405E+15 | 1,72449E+16 | 4,68766E+16 | 1,2742E+17 |
| 5 | 2,6602E+12 | 7,2313E+12 | 1,9657E+13 | 5,3432E+13 | 1,4524E+14 | 3,94815E+14 | 1,07322E+15 | 2,91731E+15 | 7,93007E+15 | 2,15562E+16 | 5,85957E+16 | 1,5928E+17 |
| 6 | 3,1923E+12 | 8,6775E+12 | 2,3588E+13 | 6,4119E+13 | 1,7429E+14 | 4,73778E+14 | 1,28786E+15 | 3,50077E+15 | 9,51608E+15 | 2,58674E+16 | 7,03149E+16 | 1,9114E+17 |
| 7 | 3,7243E+12 | 1,0124E+13 | 2,7519E+13 | 7,4805E+13 | 2,0334E+14 | 5,52741E+14 | 1,50251E+15 | 4,08423E+15 | 1,11021E+16 | 3,01786E+16 | 8,2034E+16 | 2,2299E+17 |
| 8 | 4,2564E+12 | 1,157E+13 | 3,1451E+13 | 8,5492E+13 | 2,3239E+14 | 6,31704E+14 | 1,71715E+15 | 4,66769E+15 | 1,26881E+16 | 3,44899E+16 | 9,37531E+16 | 2,5485E+17 |
| 9 | 4,7884E+12 | 1,3016E+13 | 3,5382E+13 | 9,6178E+13 | 2,6144E+14 | 7,10667E+14 | 1,93179E+15 | 5,25116E+15 | 1,42741E+16 | 3,88011E+16 | 1,05472E+17 | 2,867E+17 |
| 10 | 5,3205E+12 | 1,4463E+13 | 3,9313E+13 | 1,0686E+14 | 2,9049E+14 | 7,8963E+14 | 2,14644E+15 | 5,83462E+15 | 1,58601E+16 | 4,31123E+16 | 1,17191E+17 | 3,1856E+17 |
| 11 | 5,8525E+12 | 1,5909E+13 | 4,3245E+13 | 1,1755E+14 | 3,1954E+14 | 8,68593E+14 | 2,36108E+15 | 6,41808E+15 | 1,74461E+16 | 4,74235E+16 | 1,28911E+17 | 3,5042E+17 |
| 12 | 6,3846E+12 | 1,7355E+13 | 4,7176E+13 | 1,2824E+14 | 3,4859E+14 | 9,47556E+14 | 2,57572E+15 | 7,00154E+15 | 1,90322E+16 | 5,17348E+16 | 1,4063E+17 | 3,8227E+17 |

Values of file*e$^{col}$

| f*e$^c$ | 27 | 28 | 29 | 30 | 31 | 32 | 33 | 34 | 35 | 36 | 37 | 38 |
|---|---|---|---|---|---|---|---|---|---|---|---|---|
| 13 | 6,9166E+12 | 1,8801E+13 | 5,1107E+13 | 1,3892E+14 | 3,7764E+14 | 1,02652E+15 | 2,79037E+15 | 7,585E+15 | 2,06182E+16 | 5,6046E+16 | 1,52349E+17 | 4,1413E+17 |
| 14 | 7,4487E+12 | 2,0248E+13 | 5,5039E+13 | 1,4961E+14 | 4,0668E+14 | 1,10548E+15 | 3,00501E+15 | 8,16846E+15 | 2,22042E+16 | 6,03572E+16 | 1,64068E+17 | 4,4598E+17 |
| 15 | 7,9807E+12 | 2,1694E+13 | 5,897E+13 | 1,603E+14 | 4,3573E+14 | 1,18444E+15 | 3,21965E+15 | 8,75193E+15 | 2,37902E+16 | 6,46685E+16 | 1,75787E+17 | 4,7784E+17 |
| 16 | 8,5128E+12 | 2,314E+13 | 6,2901E+13 | 1,7098E+14 | 4,6478E+14 | 1,26341E+15 | 3,4343E+15 | 9,33539E+15 | 2,53762E+16 | 6,89797E+16 | 1,87506E+17 | 5,0969E+17 |
| 17 | 9,0448E+12 | 2,4586E+13 | 6,6833E+13 | 1,8167E+14 | 4,9383E+14 | 1,34237E+15 | 3,64894E+15 | 9,91885E+15 | 2,69622E+16 | 7,32909E+16 | 1,99225E+17 | 5,4155E+17 |
| 18 | 9,5769E+12 | 2,6033E+13 | 7,0764E+13 | 1,9236E+14 | 5,2288E+14 | 1,42133E+15 | 3,86358E+15 | 1,05023E+16 | 2,85482E+16 | 7,76022E+16 | 2,10945E+17 | 5,7341E+17 |
| 19 | 1,0109E+13 | 2,7479E+13 | 7,4695E+13 | 2,0304E+14 | 5,5193E+14 | 1,5003E+15 | 4,07823E+15 | 1,10858E+16 | 3,01343E+16 | 8,19134E+16 | 2,22664E+17 | 6,0526E+17 |
| 20 | 1,0641E+13 | 2,8925E+13 | 7,8627E+13 | 2,1373E+14 | 5,8098E+14 | 1,57926E+15 | 4,29287E+15 | 1,16692E+16 | 3,17203E+16 | 8,62246E+16 | 2,34383E+17 | 6,3712E+17 |
| 21 | 1,1173E+13 | 3,0371E+13 | 8,2558E+13 | 2,2442E+14 | 6,1003E+14 | 1,65822E+15 | 4,50752E+15 | 1,22527E+16 | 3,33063E+16 | 9,05359E+16 | 2,46102E+17 | 6,6897E+17 |
| 22 | 1,1705E+13 | 3,1818E+13 | 8,6489E+13 | 2,351E+14 | 6,3907E+14 | 1,73719E+15 | 4,72216E+15 | 1,28362E+16 | 3,48923E+16 | 9,48471E+16 | 2,57821E+17 | 7,0083E+17 |
| 23 | 1,2237E+13 | 3,3264E+13 | 9,0421E+13 | 2,4579E+14 | 6,6812E+14 | 1,81615E+15 | 4,9368E+15 | 1,34196E+16 | 3,64783E+16 | 9,91583E+16 | 2,6954E+17 | 7,3269E+17 |
| 24 | 1,2769E+13 | 3,471E+13 | 9,4352E+13 | 2,5648E+14 | 6,9717E+14 | 1,89511E+15 | 5,15145E+15 | 1,40031E+16 | 3,80643E+16 | 1,0347E+17 | 2,81259E+17 | 7,6454E+17 |
| 25 | 1,3301E+13 | 3,6156E+13 | 9,8283E+13 | 2,6716E+14 | 7,2622E+14 | 1,97407E+15 | 5,36609E+15 | 1,45865E+16 | 3,96503E+16 | 1,07781E+17 | 2,92979E+17 | 7,964E+17 |
| 26 | 1,3833E+13 | 3,7603E+13 | 1,0221E+14 | 2,7785E+14 | 7,5527E+14 | 2,05304E+15 | 5,58073E+15 | 1,517E+16 | 4,12363E+16 | 1,12092E+17 | 3,04698E+17 | 8,2825E+17 |
| 27 | 1,4365E+13 | 3,9049E+13 | 1,0615E+14 | 2,8853E+14 | 7,8432E+14 | 2,132E+15 | 5,79538E+15 | 1,57535E+16 | 4,28224E+16 | 1,16403E+17 | 3,16417E+17 | 8,6011E+17 |
| 28 | 1,4897E+13 | 4,0495E+13 | 1,1008E+14 | 2,9922E+14 | 8,1337E+14 | 2,21096E+15 | 6,01002E+15 | 1,63369E+16 | 4,44084E+16 | 1,20714E+17 | 3,28136E+17 | 8,9197E+17 |
| 29 | 1,5429E+13 | 4,1941E+13 | 1,1401E+14 | 3,0991E+14 | 8,4242E+14 | 2,28993E+15 | 6,22466E+15 | 1,69204E+16 | 4,59944E+16 | 1,25026E+17 | 3,39855E+17 | 9,2382E+17 |
| 30 | 1,5961E+13 | 4,3388E+13 | 1,1794E+14 | 3,2059E+14 | 8,7147E+14 | 2,36889E+15 | 6,43931E+15 | 1,75039E+16 | 4,75804E+16 | 1,29337E+17 | 3,51574E+17 | 9,5568E+17 |
| 31 | 1,6493E+13 | 4,4834E+13 | 1,2187E+14 | 3,3128E+14 | 9,0051E+14 | 2,44785E+15 | 6,65395E+15 | 1,80873E+16 | 4,91664E+16 | 1,33648E+17 | 3,63293E+17 | 9,8753E+17 |
| 32 | 1,7026E+13 | 4,628E+13 | 1,258E+14 | 3,4197E+14 | 9,2956E+14 | 2,52681E+15 | 6,86859E+15 | 1,86708E+16 | 5,07524E+16 | 1,37959E+17 | 3,75013E+17 | 1,0194E+18 |
| 33 | 1,7558E+13 | 4,7726E+13 | 1,2973E+14 | 3,5265E+14 | 9,5861E+14 | 2,60578E+15 | 7,08324E+15 | 1,92542E+16 | 5,23384E+16 | 1,42271E+17 | 3,86732E+17 | 1,0512E+18 |
| 34 | 1,809E+13 | 4,9173E+13 | 1,3367E+14 | 3,6334E+14 | 9,8766E+14 | 2,68474E+15 | 7,29788E+15 | 1,98377E+16 | 5,39245E+16 | 1,46582E+17 | 3,98451E+17 | 1,0831E+18 |
| 35 | 1,8622E+13 | 5,0619E+13 | 1,376E+14 | 3,7403E+14 | 1,0167E+15 | 2,7637E+15 | 7,51253E+15 | 2,04212E+16 | 5,55105E+16 | 1,50893E+17 | 4,1017E+17 | 1,115E+18 |
| 36 | 1,9154E+13 | 5,2065E+13 | 1,4153E+14 | 3,8471E+14 | 1,0458E+15 | 2,84267E+15 | 7,72717E+15 | 2,10046E+16 | 5,70965E+16 | 1,55204E+17 | 4,21889E+17 | 1,1468E+18 |
| 37 | 1,9686E+13 | 5,3512E+13 | 1,4546E+14 | 3,954E+14 | 1,0748E+15 | 2,92163E+15 | 7,94181E+15 | 2,15881E+16 | 5,86825E+16 | 1,59516E+17 | 4,33608E+17 | 1,1787E+18 |
| 38 | 2,0218E+13 | 5,4958E+13 | 1,4939E+14 | 4,0609E+14 | 1,1039E+15 | 3,00059E+15 | 8,15646E+15 | 2,21715E+16 | 6,02685E+16 | 1,63827E+17 | 4,45327E+17 | 1,2105E+18 |
| 39 | 2,075E+13 | 5,6404E+13 | 1,5332E+14 | 4,1677E+14 | 1,1329E+15 | 3,07956E+15 | 8,3711E+15 | 2,2755E+16 | 6,18545E+16 | 1,68138E+17 | 4,57047E+17 | 1,2424E+18 |
| 40 | 2,1282E+13 | 5,785E+13 | 1,5725E+14 | 4,2746E+14 | 1,162E+15 | 3,15852E+15 | 8,58574E+15 | 2,33385E+16 | 6,34405E+16 | 1,72449E+17 | 4,68766E+17 | 1,2742E+18 |
| 41 | 2,1814E+13 | 5,9297E+13 | 1,6118E+14 | 4,3815E+14 | 1,191E+15 | 3,23748E+15 | 8,80039E+15 | 2,39219E+16 | 6,50266E+16 | 1,7676E+17 | 4,80485E+17 | 1,3061E+18 |

Values of file*e$^{col}$

| f*e$^c$ | 27 | 28 | 29 | 30 | 31 | 32 | 33 | 34 | 35 | 36 | 37 | 38 |
|---|---|---|---|---|---|---|---|---|---|---|---|---|
| 42 | 2,2346E+13 | 6,0743E+13 | 1,6512E+14 | 4,4883E+14 | 1,2201E+15 | 3,31644E+15 | 9,01503E+15 | 2,45054E+16 | 6,66126E+16 | 1,81072E+17 | 4,92204E+17 | 1,3379E+18 |
| 43 | 2,2878E+13 | 6,2189E+13 | 1,6905E+14 | 4,5952E+14 | 1,2491E+15 | 3,39541E+15 | 9,22967E+15 | 2,50889E+16 | 6,81986E+16 | 1,85383E+17 | 5,03923E+17 | 1,3698E+18 |
| 44 | 2,341E+13 | 6,3635E+13 | 1,7298E+14 | 4,702E+14 | 1,2781E+15 | 3,47437E+15 | 9,44432E+15 | 2,56723E+16 | 6,97846E+16 | 1,89694E+17 | 5,15642E+17 | 1,4017E+18 |
| 45 | 2,3942E+13 | 6,5082E+13 | 1,7691E+14 | 4,8089E+14 | 1,3072E+15 | 3,55333E+15 | 9,65896E+15 | 2,62558E+16 | 7,13706E+16 | 1,94005E+17 | 5,27361E+17 | 1,4335E+18 |
| 46 | 2,4474E+13 | 6,6528E+13 | 1,8084E+14 | 4,9158E+14 | 1,3362E+15 | 3,6323E+15 | 9,8736E+15 | 2,68392E+16 | 7,29566E+16 | 1,98317E+17 | 5,39081E+17 | 1,4654E+18 |
| 47 | 2,5006E+13 | 6,7974E+13 | 1,8477E+14 | 5,0226E+14 | 1,3653E+15 | 3,71126E+15 | 1,00882E+16 | 2,74227E+16 | 7,45426E+16 | 2,02628E+17 | 5,508E+17 | 1,4972E+18 |
| 48 | 2,5538E+13 | 6,942E+13 | 1,887E+14 | 5,1295E+14 | 1,3943E+15 | 3,79022E+15 | 1,03029E+16 | 2,80062E+16 | 7,61286E+16 | 2,06939E+17 | 5,62519E+17 | 1,5291E+18 |
| 49 | 2,607E+13 | 7,0867E+13 | 1,9264E+14 | 5,2364E+14 | 1,4234E+15 | 3,86919E+15 | 1,05175E+16 | 2,85896E+16 | 7,77147E+16 | 2,1125E+17 | 5,74238E+17 | 1,5609E+18 |
| 50 | 2,6602E+13 | 7,2313E+13 | 1,9657E+14 | 5,3432E+14 | 1,4524E+15 | 3,94815E+15 | 1,07322E+16 | 2,91731E+16 | 7,93007E+16 | 2,15562E+17 | 5,85957E+17 | 1,5928E+18 |
| 51 | 2,7134E+13 | 7,3759E+13 | 2,005E+14 | 5,4501E+14 | 1,4815E+15 | 4,02711E+15 | 1,09468E+16 | 2,97565E+16 | 8,08867E+16 | 2,19873E+17 | 5,97676E+17 | 1,6247E+18 |
| 52 | 2,7667E+13 | 7,5205E+13 | 2,0443E+14 | 5,557E+14 | 1,5105E+15 | 4,10607E+15 | 1,11615E+16 | 3,034E+16 | 8,24727E+16 | 2,24184E+17 | 6,09395E+17 | 1,6565E+18 |
| 53 | 2,8199E+13 | 7,6652E+13 | 2,0836E+14 | 5,6638E+14 | 1,5396E+15 | 4,18504E+15 | 1,13761E+16 | 3,09235E+16 | 8,40587E+16 | 2,28495E+17 | 6,21115E+17 | 1,6884E+18 |
| 54 | 2,8731E+13 | 7,8098E+13 | 2,1229E+14 | 5,7707E+14 | 1,5686E+15 | 4,264E+15 | 1,15908E+16 | 3,15069E+16 | 8,56447E+16 | 2,32807E+17 | 6,32834E+17 | 1,7202E+18 |
| 55 | 2,9263E+13 | 7,9544E+13 | 2,1622E+14 | 5,8776E+14 | 1,5977E+15 | 4,34296E+15 | 1,18054E+16 | 3,20904E+16 | 8,72307E+16 | 2,37118E+17 | 6,44553E+17 | 1,7521E+18 |
| 56 | 2,9795E+13 | 8,099E+13 | 2,2015E+14 | 5,9844E+14 | 1,6267E+15 | 4,42193E+15 | 1,202E+16 | 3,26739E+16 | 8,88168E+16 | 2,41429E+17 | 6,56272E+17 | 1,7839E+18 |
| 57 | 3,0327E+13 | 8,2437E+13 | 2,2409E+14 | 6,0913E+14 | 1,6558E+15 | 4,50089E+15 | 1,22347E+16 | 3,32573E+16 | 9,04028E+16 | 2,4574E+17 | 6,67991E+17 | 1,8158E+18 |
| 58 | 3,0859E+13 | 8,3883E+13 | 2,2802E+14 | 6,1982E+14 | 1,6848E+15 | 4,57985E+15 | 1,24493E+16 | 3,38408E+16 | 9,19888E+16 | 2,50051E+17 | 6,7971E+17 | 1,8476E+18 |
| 59 | 3,1391E+13 | 8,5329E+13 | 2,3195E+14 | 6,305E+14 | 1,7139E+15 | 4,65881E+15 | 1,2664E+16 | 3,44242E+16 | 9,35748E+16 | 2,54363E+17 | 6,91429E+17 | 1,8795E+18 |
| 60 | 3,1923E+13 | 8,6775E+13 | 2,3588E+14 | 6,4119E+14 | 1,7429E+15 | 4,73778E+15 | 1,28786E+16 | 3,50077E+16 | 9,51608E+16 | 2,58674E+17 | 7,03149E+17 | 1,9114E+18 |
| 61 | 3,2455E+13 | 8,8222E+13 | 2,3981E+14 | 6,5187E+14 | 1,772E+15 | 4,81674E+15 | 1,30933E+16 | 3,55912E+16 | 9,67468E+16 | 2,62985E+17 | 7,14868E+17 | 1,9432E+18 |
| 62 | 3,2987E+13 | 8,9668E+13 | 2,4374E+14 | 6,6256E+14 | 1,801E+15 | 4,8957E+15 | 1,33079E+16 | 3,61746E+16 | 9,83328E+16 | 2,67296E+17 | 7,26587E+17 | 1,9751E+18 |
| 63 | 3,3519E+13 | 9,1114E+13 | 2,4767E+14 | 6,7325E+14 | 1,8301E+15 | 4,97467E+15 | 1,35225E+16 | 3,67581E+16 | 9,99188E+16 | 2,71608E+17 | 7,38306E+17 | 2,0069E+18 |
| 64 | 3,4051E+13 | 9,256E+13 | 2,5161E+14 | 6,8393E+14 | 1,8591E+15 | 5,05363E+15 | 1,37372E+16 | 3,73416E+16 | 1,01505E+17 | 2,75919E+17 | 7,50025E+17 | 2,0388E+18 |

Values of file*e^col

| f*e^c | 39 | 40 | 41 | 42 | 43 | 44 | 45 | 46 | 47 | 48 | 49 | 50 |
|---|---|---|---|---|---|---|---|---|---|---|---|---|
| 1 | 8,6593E+16 | 2,3539E+17 | 6,3984E+17 | 1,7393E+18 | 4,7278E+18 | 1,28516E+19 | 3,49343E+19 | 9,49612E+19 | 2,58131E+20 | 7,01674E+20 | 1,90735E+21 | 5,1847E+21 |
| 2 | 1,7319E+17 | 4,7077E+17 | 1,2797E+18 | 3,4785E+18 | 9,4557E+18 | 2,57032E+19 | 6,98685E+19 | 1,89922E+20 | 5,16263E+20 | 1,40335E+21 | 3,81469E+21 | 1,0369E+22 |
| 3 | 2,5978E+17 | 7,0616E+17 | 1,9195E+18 | 5,2178E+18 | 1,4184E+19 | 3,85548E+19 | 1,04803E+20 | 2,84884E+20 | 7,74394E+20 | 2,10502E+21 | 5,72204E+21 | 1,5554E+22 |
| 4 | 3,4637E+17 | 9,4154E+17 | 2,5594E+18 | 6,9571E+18 | 1,8911E+19 | 5,14064E+19 | 1,39737E+20 | 3,79845E+20 | 1,03253E+21 | 2,80669E+21 | 7,62939E+21 | 2,0739E+22 |
| 5 | 4,3297E+17 | 1,1769E+18 | 3,1992E+18 | 8,6964E+18 | 2,3639E+19 | 6,4258E+19 | 1,74671E+20 | 4,74806E+20 | 1,29066E+21 | 3,50837E+21 | 9,53673E+21 | 2,5924E+22 |
| 6 | 5,1956E+17 | 1,4123E+18 | 3,8391E+18 | 1,0436E+19 | 2,8367E+19 | 7,71096E+19 | 2,09606E+20 | 5,69767E+20 | 1,54879E+21 | 4,21004E+21 | 1,14441E+22 | 3,1108E+22 |
| 7 | 6,0615E+17 | 1,6477E+18 | 4,4789E+18 | 1,2175E+19 | 3,3095E+19 | 8,99612E+19 | 2,4454E+20 | 6,64728E+20 | 1,80692E+21 | 4,91172E+21 | 1,33514E+22 | 3,6293E+22 |
| 8 | 6,9275E+17 | 1,8831E+18 | 5,1187E+18 | 1,3914E+19 | 3,7823E+19 | 1,02813E+20 | 2,79474E+20 | 7,5969E+20 | 2,06505E+21 | 5,61339E+21 | 1,52588E+22 | 4,1478E+22 |
| 9 | 7,7934E+17 | 2,1185E+18 | 5,7586E+18 | 1,5653E+19 | 4,2551E+19 | 1,15664E+20 | 3,14408E+20 | 8,54651E+20 | 2,32318E+21 | 6,31506E+21 | 1,71661E+22 | 4,6662E+22 |
| 10 | 8,6593E+17 | 2,3539E+18 | 6,3984E+18 | 1,7393E+19 | 4,7278E+19 | 1,28516E+20 | 3,49343E+20 | 9,49612E+20 | 2,58131E+21 | 7,01674E+21 | 1,90735E+22 | 5,1847E+22 |
| 11 | 9,5253E+17 | 2,5892E+18 | 7,0383E+18 | 1,9132E+19 | 5,2006E+19 | 1,41368E+20 | 3,84277E+20 | 1,04457E+21 | 2,83944E+21 | 7,71841E+21 | 2,09808E+22 | 5,7032E+22 |
| 12 | 1,0391E+18 | 2,8246E+18 | 7,6781E+18 | 2,0871E+19 | 5,6734E+19 | 1,54219E+20 | 4,19211E+20 | 1,13953E+21 | 3,09758E+21 | 8,42008E+21 | 2,28882E+22 | 6,2216E+22 |
| 13 | 1,1257E+18 | 3,06E+18 | 8,318E+18 | 2,2611E+19 | 6,1462E+19 | 1,67071E+20 | 4,54146E+20 | 1,2345E+21 | 3,35571E+21 | 9,12176E+21 | 2,47955E+22 | 6,7401E+22 |
| 14 | 1,2123E+18 | 3,2954E+18 | 8,9578E+18 | 2,435E+19 | 6,619E+19 | 1,79922E+20 | 4,8908E+20 | 1,32946E+21 | 3,61384E+21 | 9,82343E+21 | 2,67029E+22 | 7,2586E+22 |
| 15 | 1,2989E+18 | 3,5308E+18 | 9,5977E+18 | 2,6089E+19 | 7,0918E+19 | 1,92774E+20 | 5,24014E+20 | 1,42442E+21 | 3,87197E+21 | 1,05251E+22 | 2,86102E+22 | 7,7771E+22 |
| 16 | 1,3855E+18 | 3,7662E+18 | 1,0237E+19 | 2,7828E+19 | 7,5645E+19 | 2,05626E+20 | 5,58948E+20 | 1,51938E+21 | 4,1301E+21 | 1,12268E+22 | 3,05175E+22 | 8,2955E+22 |
| 17 | 1,4721E+18 | 4,0015E+18 | 1,0877E+19 | 2,9568E+19 | 8,0373E+19 | 2,18477E+20 | 5,93883E+20 | 1,61434E+21 | 4,38823E+21 | 1,19285E+22 | 3,24249E+22 | 8,814E+22 |
| 18 | 1,5587E+18 | 4,2369E+18 | 1,1517E+19 | 3,1307E+19 | 8,5101E+19 | 2,31329E+20 | 6,28817E+20 | 1,7093E+21 | 4,64636E+21 | 1,26301E+22 | 3,43322E+22 | 9,3325E+22 |
| 19 | 1,6453E+18 | 4,4723E+18 | 1,2157E+19 | 3,3046E+19 | 8,9829E+19 | 2,4418E+20 | 6,63751E+20 | 1,80426E+21 | 4,90449E+21 | 1,33318E+22 | 3,62396E+22 | 9,8509E+22 |
| 20 | 1,7319E+18 | 4,7077E+18 | 1,2797E+19 | 3,4785E+19 | 9,4557E+19 | 2,57032E+20 | 6,98685E+20 | 1,89922E+21 | 5,16263E+21 | 1,40335E+22 | 3,81469E+22 | 1,0369E+23 |
| 21 | 1,8185E+18 | 4,9431E+18 | 1,3437E+19 | 3,6525E+19 | 9,9285E+19 | 2,69884E+20 | 7,3362E+20 | 1,99419E+21 | 5,42076E+21 | 1,47351E+22 | 4,00543E+22 | 1,0888E+23 |
| 22 | 1,9051E+18 | 5,1785E+18 | 1,4077E+19 | 3,8264E+19 | 1,0401E+20 | 2,82735E+20 | 7,68554E+20 | 2,08915E+21 | 5,67889E+21 | 1,54368E+22 | 4,19616E+22 | 1,1406E+23 |
| 23 | 1,9916E+18 | 5,4139E+18 | 1,4716E+19 | 4,0003E+19 | 1,0874E+20 | 2,95587E+20 | 8,03488E+20 | 2,18411E+21 | 5,93702E+21 | 1,61385E+22 | 4,3869E+22 | 1,1925E+23 |
| 24 | 2,0782E+18 | 5,6492E+18 | 1,5356E+19 | 4,1743E+19 | 1,1347E+20 | 3,08438E+20 | 8,38423E+20 | 2,27907E+21 | 6,19515E+21 | 1,68402E+22 | 4,57763E+22 | 1,2443E+23 |
| 25 | 2,1648E+18 | 5,8846E+18 | 1,5996E+19 | 4,3482E+19 | 1,182E+20 | 3,2129E+20 | 8,73357E+20 | 2,37403E+21 | 6,45328E+21 | 1,75418E+22 | 4,76837E+22 | 1,2962E+23 |
| 26 | 2,2514E+18 | 6,12E+18 | 1,6636E+19 | 4,5221E+19 | 1,2292E+20 | 3,34142E+20 | 9,08291E+20 | 2,46899E+21 | 6,71141E+21 | 1,82435E+22 | 4,9591E+22 | 1,348E+23 |
| 27 | 2,338E+18 | 6,3554E+18 | 1,7276E+19 | 4,696E+19 | 1,2765E+20 | 3,46993E+20 | 9,43225E+20 | 2,56395E+21 | 6,96954E+21 | 1,89452E+22 | 5,14984E+22 | 1,3999E+23 |
| 28 | 2,4246E+18 | 6,5908E+18 | 1,7916E+19 | 4,87E+19 | 1,3238E+20 | 3,59845E+20 | 9,7816E+20 | 2,65891E+21 | 7,22768E+21 | 1,96469E+22 | 5,34057E+22 | 1,4517E+23 |
| 29 | 2,5112E+18 | 6,8262E+18 | 1,8555E+19 | 5,0439E+19 | 1,3711E+20 | 3,72696E+20 | 1,01309E+21 | 2,75387E+21 | 7,48581E+21 | 2,03485E+22 | 5,53131E+22 | 1,5036E+23 |

Values of file*e$^{col}$

| f*e$^c$ | 39 | 40 | 41 | 42 | 43 | 44 | 45 | 46 | 47 | 48 | 49 | 50 |
|---|---|---|---|---|---|---|---|---|---|---|---|---|
| 30 | 2,5978E+18 | 7,0616E+18 | 1,9195E+19 | 5,2178E+19 | 1,4184E+20 | 3,85548E+20 | 1,04803E+21 | 2,84884E+21 | 7,74394E+21 | 2,10502E+22 | 5,72204E+22 | 1,5554E+23 |
| 31 | 2,6844E+18 | 7,2969E+18 | 1,9835E+19 | 5,3918E+19 | 1,4656E+20 | 3,984E+20 | 1,08296E+21 | 2,9438E+21 | 8,00207E+21 | 2,17519E+22 | 5,91277E+22 | 1,6073E+23 |
| 32 | 2,771E+18 | 7,5323E+18 | 2,0475E+19 | 5,5657E+19 | 1,5129E+20 | 4,11251E+20 | 1,1179E+21 | 3,03876E+21 | 8,2602E+21 | 2,24536E+22 | 6,10351E+22 | 1,6591E+23 |
| 33 | 2,8576E+18 | 7,7677E+18 | 2,1115E+19 | 5,7396E+19 | 1,5602E+20 | 4,24103E+20 | 1,15283E+21 | 3,13372E+21 | 8,51833E+21 | 2,31552E+22 | 6,29424E+22 | 1,711E+23 |
| 34 | 2,9442E+18 | 8,0031E+18 | 2,1755E+19 | 5,9135E+19 | 1,6075E+20 | 4,36954E+20 | 1,18777E+21 | 3,22868E+21 | 8,77646E+21 | 2,38569E+22 | 6,48498E+22 | 1,7628E+23 |
| 35 | 3,0308E+18 | 8,2385E+18 | 2,2395E+19 | 6,0875E+19 | 1,6547E+20 | 4,49806E+20 | 1,2227E+21 | 3,32364E+21 | 9,0346E+21 | 2,45586E+22 | 6,67571E+22 | 1,8146E+23 |
| 36 | 3,1174E+18 | 8,4739E+18 | 2,3034E+19 | 6,2614E+19 | 1,702E+20 | 4,62658E+20 | 1,25763E+21 | 3,4186E+21 | 9,29273E+21 | 2,52602E+22 | 6,86645E+22 | 1,8665E+23 |
| 37 | 3,204E+18 | 8,7093E+18 | 2,3674E+19 | 6,4353E+19 | 1,7493E+20 | 4,75509E+20 | 1,29257E+21 | 3,51356E+21 | 9,55086E+21 | 2,59619E+22 | 7,05718E+22 | 1,9183E+23 |
| 38 | 3,2905E+18 | 8,9446E+18 | 2,4314E+19 | 6,6092E+19 | 1,7966E+20 | 4,88361E+20 | 1,3275E+21 | 3,60853E+21 | 9,80899E+21 | 2,66636E+22 | 7,24792E+22 | 1,9702E+23 |
| 39 | 3,3771E+18 | 9,18E+18 | 2,4954E+19 | 6,7832E+19 | 1,8439E+20 | 5,01212E+20 | 1,36244E+21 | 3,70349E+21 | 1,00671E+22 | 2,73653E+22 | 7,43865E+22 | 2,022E+23 |
| 40 | 3,4637E+18 | 9,4154E+18 | 2,5594E+19 | 6,9571E+19 | 1,8911E+20 | 5,14064E+20 | 1,39737E+21 | 3,79845E+21 | 1,03253E+22 | 2,80669E+22 | 7,62939E+22 | 2,0739E+23 |
| 41 | 3,5503E+18 | 9,6508E+18 | 2,6234E+19 | 7,131E+19 | 1,9384E+20 | 5,26916E+20 | 1,43231E+21 | 3,89341E+21 | 1,05834E+22 | 2,87686E+22 | 7,82012E+22 | 2,1257E+23 |
| 42 | 3,6369E+18 | 9,8862E+18 | 2,6873E+19 | 7,305E+19 | 1,9857E+20 | 5,39767E+20 | 1,46724E+21 | 3,98837E+21 | 1,08415E+22 | 2,94703E+22 | 8,01086E+22 | 2,1776E+23 |
| 43 | 3,7235E+18 | 1,0122E+19 | 2,7513E+19 | 7,4789E+19 | 2,033E+20 | 5,52619E+20 | 1,50217E+21 | 4,08333E+21 | 1,10996E+22 | 3,0172E+22 | 8,20159E+22 | 2,2294E+23 |
| 44 | 3,8101E+18 | 1,0357E+19 | 2,8153E+19 | 7,6528E+19 | 2,0802E+20 | 5,6547E+20 | 1,53711E+21 | 4,17829E+21 | 1,13578E+22 | 3,08736E+22 | 8,39232E+22 | 2,2813E+23 |
| 45 | 3,8967E+18 | 1,0592E+19 | 2,8793E+19 | 7,8267E+19 | 2,1275E+20 | 5,78322E+20 | 1,57204E+21 | 4,27325E+21 | 1,16159E+22 | 3,15753E+22 | 8,58306E+22 | 2,3331E+23 |
| 46 | 3,9833E+18 | 1,0828E+19 | 2,9433E+19 | 8,0007E+19 | 2,1748E+20 | 5,91174E+20 | 1,60698E+21 | 4,36821E+21 | 1,1874E+22 | 3,2277E+22 | 8,77379E+22 | 2,385E+23 |
| 47 | 4,0699E+18 | 1,1063E+19 | 3,0073E+19 | 8,1746E+19 | 2,2221E+20 | 6,04025E+20 | 1,64191E+21 | 4,46318E+21 | 1,21322E+22 | 3,29787E+22 | 8,96453E+22 | 2,4368E+23 |
| 48 | 4,1565E+18 | 1,1298E+19 | 3,0712E+19 | 8,3485E+19 | 2,2694E+20 | 6,16877E+20 | 1,67685E+21 | 4,55814E+21 | 1,23903E+22 | 3,36803E+22 | 9,15526E+22 | 2,4887E+23 |
| 49 | 4,2431E+18 | 1,1534E+19 | 3,1352E+19 | 8,5224E+19 | 2,3166E+20 | 6,29728E+20 | 1,71178E+21 | 4,6531E+21 | 1,26484E+22 | 3,4382E+22 | 9,346E+22 | 2,5405E+23 |
| 50 | 4,3297E+18 | 1,1769E+19 | 3,1992E+19 | 8,6964E+19 | 2,3639E+20 | 6,4258E+20 | 1,74671E+21 | 4,74806E+21 | 1,29066E+22 | 3,50837E+22 | 9,53673E+22 | 2,5924E+23 |
| 51 | 4,4163E+18 | 1,2005E+19 | 3,2632E+19 | 8,8703E+19 | 2,4112E+20 | 6,55432E+20 | 1,78165E+21 | 4,84302E+21 | 1,31647E+22 | 3,57854E+22 | 9,72747E+22 | 2,6442E+23 |
| 52 | 4,5029E+18 | 1,224E+19 | 3,3272E+19 | 9,0442E+19 | 2,4585E+20 | 6,68283E+20 | 1,81658E+21 | 4,93798E+21 | 1,34228E+22 | 3,6487E+22 | 9,9182E+22 | 2,696E+23 |
| 53 | 4,5895E+18 | 1,2475E+19 | 3,3912E+19 | 9,2182E+19 | 2,5058E+20 | 6,81135E+20 | 1,85152E+21 | 5,03294E+21 | 1,3681E+22 | 3,71887E+22 | 1,01089E+23 | 2,7479E+23 |
| 54 | 4,676E+18 | 1,2711E+19 | 3,4552E+19 | 9,3921E+19 | 2,553E+20 | 6,93986E+20 | 1,88645E+21 | 5,1279E+21 | 1,39391E+22 | 3,78904E+22 | 1,02997E+23 | 2,7997E+23 |
| 55 | 4,7626E+18 | 1,2946E+19 | 3,5191E+19 | 9,566E+19 | 2,6003E+20 | 7,06838E+20 | 1,92138E+21 | 5,22287E+21 | 1,41972E+22 | 3,8592E+22 | 1,04904E+23 | 2,8516E+23 |
| 56 | 4,8492E+18 | 1,3182E+19 | 3,5831E+19 | 9,7399E+19 | 2,6476E+20 | 7,1969E+20 | 1,95632E+21 | 5,31783E+21 | 1,44554E+22 | 3,92937E+22 | 1,06811E+23 | 2,9034E+23 |
| 57 | 4,9358E+18 | 1,3417E+19 | 3,6471E+19 | 9,9139E+19 | 2,6949E+20 | 7,32541E+20 | 1,99125E+21 | 5,41279E+21 | 1,47135E+22 | 3,99954E+22 | 1,08719E+23 | 2,9553E+23 |
| 58 | 5,0224E+18 | 1,3652E+19 | 3,7111E+19 | 1,0088E+20 | 2,7421E+20 | 7,45393E+20 | 2,02619E+21 | 5,50775E+21 | 1,49716E+22 | 4,06971E+22 | 1,10626E+23 | 3,0071E+23 |

Values of file*e<sup>col</sup>

| f*e<sup>c</sup> | 39 | 40 | 41 | 42 | 43 | 44 | 45 | 46 | 47 | 48 | 49 | 50 |
|---|---|---|---|---|---|---|---|---|---|---|---|---|
| 59 | 5,109E+18 | 1,3888E+19 | 3,7751E+19 | 1,0262E+20 | 2,7894E+20 | 7,58244E+20 | 2,06112E+21 | 5,60271E+21 | 1,52297E+22 | 4,13987E+22 | 1,12533E+23 | 3,059E+23 |
| 60 | 5,1956E+18 | 1,4123E+19 | 3,8391E+19 | 1,0436E+20 | 2,8367E+20 | 7,71096E+20 | 2,09606E+21 | 5,69767E+21 | 1,54879E+22 | 4,21004E+22 | 1,14441E+23 | 3,1108E+23 |
| 61 | 5,2822E+18 | 1,4359E+19 | 3,903E+19 | 1,061E+20 | 2,884E+20 | 7,83948E+20 | 2,13099E+21 | 5,79263E+21 | 1,5746E+22 | 4,28021E+22 | 1,16348E+23 | 3,1627E+23 |
| 62 | 5,3688E+18 | 1,4594E+19 | 3,967E+19 | 1,0784E+20 | 2,9313E+20 | 7,96799E+20 | 2,16592E+21 | 5,88759E+21 | 1,60041E+22 | 4,35038E+22 | 1,18255E+23 | 3,2145E+23 |
| 63 | 5,4554E+18 | 1,4829E+19 | 4,031E+19 | 1,0957E+20 | 2,9785E+20 | 8,09651E+20 | 2,20086E+21 | 5,98256E+21 | 1,62623E+22 | 4,42054E+22 | 1,20163E+23 | 3,2664E+23 |
| 64 | 5,542E+18 | 1,5065E+19 | 4,095E+19 | 1,1131E+20 | 3,0258E+20 | 8,22502E+20 | 2,23579E+21 | 6,07752E+21 | 1,65204E+22 | 4,49071E+22 | 1,2207E+23 | 3,3182E+23 |

Values of file*e$^{col}$

| f*e$^c$ | 51 | 52 | 53 | 54 | 55 | 56 | 57 | 58 |
|---|---|---|---|---|---|---|---|---|
| 1 | 1,4093E+22 | 3,831E+22 | 1,0414E+23 | 2,8308E+23 | 7,6948E+23 | 2,09166E+24 | 5,68572E+24 | 1,54554E+25 |
| 2 | 2,8187E+22 | 7,662E+22 | 2,0828E+23 | 5,6615E+23 | 1,539E+24 | 4,18332E+24 | 1,13714E+25 | 3,09108E+25 |
| 3 | 4,228E+22 | 1,1493E+23 | 3,1241E+23 | 8,4923E+23 | 2,3084E+24 | 6,27498E+24 | 1,70572E+25 | 4,63662E+25 |
| 4 | 5,6374E+22 | 1,5324E+23 | 4,1655E+23 | 1,1323E+24 | 3,0779E+24 | 8,36664E+24 | 2,27429E+25 | 6,18216E+25 |
| 5 | 7,0467E+22 | 1,9155E+23 | 5,2069E+23 | 1,4154E+24 | 3,8474E+24 | 1,04583E+25 | 2,84286E+25 | 7,72769E+25 |
| 6 | 8,4561E+22 | 2,2986E+23 | 6,2483E+23 | 1,6985E+24 | 4,6169E+24 | 1,255E+25 | 3,41143E+25 | 9,27323E+25 |
| 7 | 9,8654E+22 | 2,6817E+23 | 7,2896E+23 | 1,9815E+24 | 5,3863E+24 | 1,46416E+25 | 3,98E+25 | 1,08188E+26 |
| 8 | 1,1275E+23 | 3,0648E+23 | 8,331E+23 | 2,2646E+24 | 6,1558E+24 | 1,67333E+25 | 4,54858E+25 | 1,23643E+26 |
| 9 | 1,2684E+23 | 3,4479E+23 | 9,3724E+23 | 2,5477E+24 | 6,9253E+24 | 1,88249E+25 | 5,11715E+25 | 1,39099E+26 |
| 10 | 1,4093E+23 | 3,831E+23 | 1,0414E+24 | 2,8308E+24 | 7,6948E+24 | 2,09166E+25 | 5,68572E+25 | 1,54554E+26 |
| 11 | 1,5503E+23 | 4,2141E+23 | 1,1455E+24 | 3,1138E+24 | 8,4643E+24 | 2,30083E+25 | 6,25429E+25 | 1,70009E+26 |
| 12 | 1,6912E+23 | 4,5972E+23 | 1,2497E+24 | 3,3969E+24 | 9,2337E+24 | 2,50999E+25 | 6,82286E+25 | 1,85465E+26 |
| 13 | 1,8322E+23 | 4,9803E+23 | 1,3538E+24 | 3,68E+24 | 1,0003E+25 | 2,71916E+25 | 7,39144E+25 | 2,0092E+26 |
| 14 | 1,9731E+23 | 5,3634E+23 | 1,4579E+24 | 3,9631E+24 | 1,0773E+25 | 2,92832E+25 | 7,96001E+25 | 2,16375E+26 |
| 15 | 2,114E+23 | 5,7465E+23 | 1,5621E+24 | 4,2461E+24 | 1,1542E+25 | 3,13749E+25 | 8,52858E+25 | 2,31831E+26 |
| 16 | 2,255E+23 | 6,1296E+23 | 1,6662E+24 | 4,5292E+24 | 1,2312E+25 | 3,34666E+25 | 9,09715E+25 | 2,47286E+26 |
| 17 | 2,3959E+23 | 6,5127E+23 | 1,7703E+24 | 4,8123E+24 | 1,3081E+25 | 3,55582E+25 | 9,66572E+25 | 2,62742E+26 |
| 18 | 2,5368E+23 | 6,8958E+23 | 1,8745E+24 | 5,0954E+24 | 1,3851E+25 | 3,76499E+25 | 1,02343E+26 | 2,78197E+26 |
| 19 | 2,6778E+23 | 7,2789E+23 | 1,9786E+24 | 5,3784E+24 | 1,462E+25 | 3,97415E+25 | 1,08029E+26 | 2,93652E+26 |
| 20 | 2,8187E+23 | 7,662E+23 | 2,0828E+24 | 5,6615E+24 | 1,539E+25 | 4,18332E+25 | 1,13714E+26 | 3,09108E+26 |
| 21 | 2,9596E+23 | 8,0451E+23 | 2,1869E+24 | 5,9446E+24 | 1,6159E+25 | 4,39248E+25 | 1,194E+26 | 3,24563E+26 |
| 22 | 3,1006E+23 | 8,4282E+23 | 2,291E+24 | 6,2277E+24 | 1,6929E+25 | 4,60165E+25 | 1,25086E+26 | 3,40019E+26 |
| 23 | 3,2415E+23 | 8,8113E+23 | 2,3952E+24 | 6,5107E+24 | 1,7698E+25 | 4,81082E+25 | 1,30772E+26 | 3,55474E+26 |
| 24 | 3,3824E+23 | 9,1944E+23 | 2,4993E+24 | 6,7938E+24 | 1,8467E+25 | 5,01998E+25 | 1,36457E+26 | 3,70929E+26 |
| 25 | 3,5234E+23 | 9,5775E+23 | 2,6034E+24 | 7,0769E+24 | 1,9237E+25 | 5,22915E+25 | 1,42143E+26 | 3,86385E+26 |
| 26 | 3,6643E+23 | 9,9606E+23 | 2,7076E+24 | 7,36E+24 | 2,0006E+25 | 5,43831E+25 | 1,47829E+26 | 4,0184E+26 |
| 27 | 3,8052E+23 | 1,0344E+24 | 2,8117E+24 | 7,643E+24 | 2,0776E+25 | 5,64748E+25 | 1,53514E+26 | 4,17296E+26 |
| 28 | 3,9462E+23 | 1,0727E+24 | 2,9159E+24 | 7,9261E+24 | 2,1545E+25 | 5,85665E+25 | 1,592E+26 | 4,32751E+26 |
| 29 | 4,0871E+23 | 1,111E+24 | 3,02E+24 | 8,2092E+24 | 2,2315E+25 | 6,06581E+25 | 1,64886E+26 | 4,48206E+26 |
| 30 | 4,228E+23 | 1,1493E+24 | 3,1241E+24 | 8,4923E+24 | 2,3084E+25 | 6,27498E+25 | 1,70572E+26 | 4,63662E+26 |
| 31 | 4,369E+23 | 1,1876E+24 | 3,2283E+24 | 8,7753E+24 | 2,3854E+25 | 6,48414E+25 | 1,76257E+26 | 4,79117E+26 |
| 32 | 4,5099E+23 | 1,2259E+24 | 3,3324E+24 | 9,0584E+24 | 2,4623E+25 | 6,69331E+25 | 1,81943E+26 | 4,94572E+26 |
| 33 | 4,6509E+23 | 1,2642E+24 | 3,4365E+24 | 9,3415E+24 | 2,5393E+25 | 6,90248E+25 | 1,87629E+26 | 5,10028E+26 |
| 34 | 4,7918E+23 | 1,3025E+24 | 3,5407E+24 | 9,6246E+24 | 2,6162E+25 | 7,11164E+25 | 1,93314E+26 | 5,25483E+26 |
| 35 | 4,9327E+23 | 1,3409E+24 | 3,6448E+24 | 9,9076E+24 | 2,6932E+25 | 7,32081E+25 | 1,99E+26 | 5,40939E+26 |
| 36 | 5,0737E+23 | 1,3792E+24 | 3,749E+24 | 1,0191E+25 | 2,7701E+25 | 7,52997E+25 | 2,04686E+26 | 5,56394E+26 |
| 37 | 5,2146E+23 | 1,4175E+24 | 3,8531E+24 | 1,0474E+25 | 2,8471E+25 | 7,73914E+25 | 2,10372E+26 | 5,71849E+26 |
| 38 | 5,3555E+23 | 1,4558E+24 | 3,9572E+24 | 1,0757E+25 | 2,924E+25 | 7,94831E+25 | 2,16057E+26 | 5,87305E+26 |
| 39 | 5,4965E+23 | 1,4941E+24 | 4,0614E+24 | 1,104E+25 | 3,001E+25 | 8,15747E+25 | 2,21743E+26 | 6,0276E+26 |
| 40 | 5,6374E+23 | 1,5324E+24 | 4,1655E+24 | 1,1323E+25 | 3,0779E+25 | 8,36664E+25 | 2,27429E+26 | 6,18216E+26 |
| 41 | 5,7783E+23 | 1,5707E+24 | 4,2696E+24 | 1,1606E+25 | 3,1549E+25 | 8,5758E+25 | 2,33115E+26 | 6,33671E+26 |
| 42 | 5,9193E+23 | 1,609E+24 | 4,3738E+24 | 1,1889E+25 | 3,2318E+25 | 8,78497E+25 | 2,388E+26 | 6,49126E+26 |
| 43 | 6,0602E+23 | 1,6473E+24 | 4,4779E+24 | 1,2172E+25 | 3,3088E+25 | 8,99414E+25 | 2,44486E+26 | 6,64582E+26 |
| 44 | 6,2011E+23 | 1,6856E+24 | 4,5821E+24 | 1,2455E+25 | 3,3857E+25 | 9,2033E+25 | 2,50172E+26 | 6,80037E+26 |
| 45 | 6,3421E+23 | 1,724E+24 | 4,6862E+24 | 1,2738E+25 | 3,4627E+25 | 9,41247E+25 | 2,55857E+26 | 6,95493E+26 |

Values of file*e^col

| | | | | | | | | |
|---|---|---|---|---|---|---|---|---|
| 46 | 6,483E+23 | 1,7623E+24 | 4,7903E+24 | 1,3021E+25 | 3,5396E+25 | 9,62163E+25 | 2,61543E+26 | 7,10948E+26 |
| f*e^c | 51 | 52 | 53 | 54 | 55 | 56 | 57 | 58 |
| 47 | 6,6239E+23 | 1,8006E+24 | 4,8945E+24 | 1,3305E+25 | 3,6165E+25 | 9,8308E+25 | 2,67229E+26 | 7,26403E+26 |
| 48 | 6,7649E+23 | 1,8389E+24 | 4,9986E+24 | 1,3588E+25 | 3,6935E+25 | 1,004E+26 | 2,72915E+26 | 7,41859E+26 |
| 49 | 6,9058E+23 | 1,8772E+24 | 5,1027E+24 | 1,3871E+25 | 3,7704E+25 | 1,02491E+26 | 2,786E+26 | 7,57314E+26 |
| 50 | 7,0467E+23 | 1,9155E+24 | 5,2069E+24 | 1,4154E+25 | 3,8474E+25 | 1,04583E+26 | 2,84286E+26 | 7,72769E+26 |
| 51 | 7,1877E+23 | 1,9538E+24 | 5,311E+24 | 1,4437E+25 | 3,9243E+25 | 1,06675E+26 | 2,89972E+26 | 7,88225E+26 |
| 52 | 7,3286E+23 | 1,9921E+24 | 5,4152E+24 | 1,472E+25 | 4,0013E+25 | 1,08766E+26 | 2,95657E+26 | 8,0368E+26 |
| 53 | 7,4696E+23 | 2,0304E+24 | 5,5193E+24 | 1,5003E+25 | 4,0782E+25 | 1,10858E+26 | 3,01343E+26 | 8,19136E+26 |
| 54 | 7,6105E+23 | 2,0687E+24 | 5,6234E+24 | 1,5286E+25 | 4,1552E+25 | 1,1295E+26 | 3,07029E+26 | 8,34591E+26 |
| 55 | 7,7514E+23 | 2,1071E+24 | 5,7276E+24 | 1,5569E+25 | 4,2321E+25 | 1,15041E+26 | 3,12715E+26 | 8,50046E+26 |
| 56 | 7,8924E+23 | 2,1454E+24 | 5,8317E+24 | 1,5852E+25 | 4,3091E+25 | 1,17133E+26 | 3,184E+26 | 8,65502E+26 |
| 57 | 8,0333E+23 | 2,1837E+24 | 5,9358E+24 | 1,6135E+25 | 4,386E+25 | 1,19225E+26 | 3,24086E+26 | 8,80957E+26 |
| 58 | 8,1742E+23 | 2,222E+24 | 6,04E+24 | 1,6418E+25 | 4,463E+25 | 1,21316E+26 | 3,29772E+26 | 8,96413E+26 |
| 59 | 8,3152E+23 | 2,2603E+24 | 6,1441E+24 | 1,6701E+25 | 4,5399E+25 | 1,23408E+26 | 3,35457E+26 | 9,11868E+26 |
| 60 | 8,4561E+23 | 2,2986E+24 | 6,2483E+24 | 1,6985E+25 | 4,6169E+25 | 1,255E+26 | 3,41143E+26 | 9,27323E+26 |
| 61 | 8,597E+23 | 2,3369E+24 | 6,3524E+24 | 1,7268E+25 | 4,6938E+25 | 1,27591E+26 | 3,46829E+26 | 9,42779E+26 |
| 62 | 8,738E+23 | 2,3752E+24 | 6,4565E+24 | 1,7551E+25 | 4,7708E+25 | 1,29683E+26 | 3,52515E+26 | 9,58234E+26 |
| 63 | 8,8789E+23 | 2,4135E+24 | 6,5607E+24 | 1,7834E+25 | 4,8477E+25 | 1,31775E+26 | 3,582E+26 | 9,7369E+26 |
| 64 | 9,0198E+23 | 2,4518E+24 | 6,6648E+24 | 1,8117E+25 | 4,9247E+25 | 1,33866E+26 | 3,63886E+26 | 9,89145E+26 |

| f*e^c | 59 | 60 | 61 | 62 | 63 | 64 | 65 |
|---|---|---|---|---|---|---|---|
| 1 | 4,2012E+25 | 1,142E+26 | 3,1043E+26 | 8,4384E+26 | 2,2938E+27 | 6,23515E+27 | 1,69489E+28 |
| 2 | 8,4024E+25 | 2,284E+26 | 6,2086E+26 | 1,6877E+27 | 4,5876E+27 | 1,24703E+28 | 3,38978E+28 |
| 3 | 1,2604E+26 | 3,426E+26 | 9,3129E+26 | 2,5315E+27 | 6,8813E+27 | 1,87054E+28 | 5,08467E+28 |
| 4 | 1,6805E+26 | 4,568E+26 | 1,2417E+27 | 3,3753E+27 | 9,1751E+27 | 2,49406E+28 | 6,77956E+28 |
| 5 | 2,1006E+26 | 5,71E+26 | 1,5521E+27 | 4,2192E+27 | 1,1469E+28 | 3,11757E+28 | 8,47445E+28 |
| 6 | 2,5207E+26 | 6,852E+26 | 1,8626E+27 | 5,063E+27 | 1,3763E+28 | 3,74109E+28 | 1,01693E+29 |
| 7 | 2,9408E+26 | 7,9941E+26 | 2,173E+27 | 5,9068E+27 | 1,6056E+28 | 4,3646E+28 | 1,18642E+29 |
| 8 | 3,361E+26 | 9,1361E+26 | 2,4834E+27 | 6,7507E+27 | 1,835E+28 | 4,98812E+28 | 1,35591E+29 |
| 9 | 3,7811E+26 | 1,0278E+27 | 2,7939E+27 | 7,5945E+27 | 2,0644E+28 | 5,61163E+28 | 1,5254E+29 |
| 10 | 4,2012E+26 | 1,142E+27 | 3,1043E+27 | 8,4384E+27 | 2,2938E+28 | 6,23515E+28 | 1,69489E+29 |
| 11 | 4,6213E+26 | 1,2562E+27 | 3,4147E+27 | 9,2822E+27 | 2,5232E+28 | 6,85866E+28 | 1,86438E+29 |
| 12 | 5,0415E+26 | 1,3704E+27 | 3,7252E+27 | 1,0126E+28 | 2,7525E+28 | 7,48218E+28 | 2,03387E+29 |
| 13 | 5,4616E+26 | 1,4846E+27 | 4,0356E+27 | 1,097E+28 | 2,9819E+28 | 8,10569E+28 | 2,20336E+29 |
| 14 | 5,8817E+26 | 1,5988E+27 | 4,346E+27 | 1,1814E+28 | 3,2113E+28 | 8,72921E+28 | 2,37284E+29 |
| 15 | 6,3018E+26 | 1,713E+27 | 4,6564E+27 | 1,2658E+28 | 3,4407E+28 | 9,35272E+28 | 2,54233E+29 |
| 16 | 6,7219E+26 | 1,8272E+27 | 4,9669E+27 | 1,3501E+28 | 3,6701E+28 | 9,97624E+28 | 2,71182E+29 |
| 17 | 7,1421E+26 | 1,9414E+27 | 5,2773E+27 | 1,4345E+28 | 3,8994E+28 | 1,05998E+29 | 2,88131E+29 |
| 18 | 7,5622E+26 | 2,0556E+27 | 5,5877E+27 | 1,5189E+28 | 4,1288E+28 | 1,12233E+29 | 3,0508E+29 |
| 19 | 7,9823E+26 | 2,1698E+27 | 5,8982E+27 | 1,6033E+28 | 4,3582E+28 | 1,18468E+29 | 3,22029E+29 |
| 20 | 8,4024E+26 | 2,284E+27 | 6,2086E+27 | 1,6877E+28 | 4,5876E+28 | 1,24703E+29 | 3,38978E+29 |
| 21 | 8,8225E+26 | 2,3982E+27 | 6,519E+27 | 1,7721E+28 | 4,8169E+28 | 1,30938E+29 | 3,55927E+29 |
| 22 | 9,2427E+26 | 2,5124E+27 | 6,8295E+27 | 1,8564E+28 | 5,0463E+28 | 1,37173E+29 | 3,72876E+29 |
| 23 | 9,6628E+26 | 2,6266E+27 | 7,1399E+27 | 1,9408E+28 | 5,2757E+28 | 1,43408E+29 | 3,89825E+29 |
| 24 | 1,0083E+27 | 2,7408E+27 | 7,4503E+27 | 2,0252E+28 | 5,5051E+28 | 1,49644E+29 | 4,06773E+29 |
| 25 | 1,0503E+27 | 2,855E+27 | 7,7607E+27 | 2,1096E+28 | 5,7345E+28 | 1,55879E+29 | 4,23722E+29 |

Values of file*e^col

| | | | | | | | |
|---|---|---|---|---|---|---|---|
| 26 | 1,0923E+27 | 2,9692E+27 | 8,0712E+27 | 2,194E+28 | 5,9638E+28 | 1,62114E+29 | 4,40671E+29 |
| 27 | 1,1343E+27 | 3,0834E+27 | 8,3816E+27 | 2,2784E+28 | 6,1932E+28 | 1,68349E+29 | 4,5762E+29 |
| f*e^c | 59 | 60 | 61 | 62 | 63 | 64 | 65 |
| 28 | 1,1763E+27 | 3,1976E+27 | 8,692E+27 | 2,3627E+28 | 6,4226E+28 | 1,74584E+29 | 4,74569E+29 |
| 29 | 1,2184E+27 | 3,3118E+27 | 9,0025E+27 | 2,4471E+28 | 6,652E+28 | 1,80819E+29 | 4,91518E+29 |
| 30 | 1,2604E+27 | 3,426E+27 | 9,3129E+27 | 2,5315E+28 | 6,8813E+28 | 1,87054E+29 | 5,08467E+29 |
| 31 | 1,3024E+27 | 3,5402E+27 | 9,6233E+27 | 2,6159E+28 | 7,1107E+28 | 1,9329E+29 | 5,25416E+29 |
| 32 | 1,3444E+27 | 3,6544E+27 | 9,9338E+27 | 2,7003E+28 | 7,3401E+28 | 1,99525E+29 | 5,42365E+29 |
| 33 | 1,3864E+27 | 3,7686E+27 | 1,0244E+28 | 2,7847E+28 | 7,5695E+28 | 2,0576E+29 | 5,59313E+29 |
| 34 | 1,4284E+27 | 3,8828E+27 | 1,0555E+28 | 2,869E+28 | 7,7989E+28 | 2,11995E+29 | 5,76262E+29 |
| 35 | 1,4704E+27 | 3,997E+27 | 1,0865E+28 | 2,9534E+28 | 8,0282E+28 | 2,1823E+29 | 5,93211E+29 |
| 36 | 1,5124E+27 | 4,1112E+27 | 1,1175E+28 | 3,0378E+28 | 8,2576E+28 | 2,24465E+29 | 6,1016E+29 |
| 37 | 1,5544E+27 | 4,2254E+27 | 1,1486E+28 | 3,1222E+28 | 8,487E+28 | 2,30701E+29 | 6,27109E+29 |
| 38 | 1,5965E+27 | 4,3396E+27 | 1,1796E+28 | 3,2066E+28 | 8,7164E+28 | 2,36936E+29 | 6,44058E+29 |
| 39 | 1,6385E+27 | 4,4538E+27 | 1,2107E+28 | 3,291E+28 | 8,9458E+28 | 2,43171E+29 | 6,61007E+29 |
| 40 | 1,6805E+27 | 4,568E+27 | 1,2417E+28 | 3,3753E+28 | 9,1751E+28 | 2,49406E+29 | 6,77956E+29 |
| 41 | 1,7225E+27 | 4,6822E+27 | 1,2728E+28 | 3,4597E+28 | 9,4045E+28 | 2,55641E+29 | 6,94905E+29 |
| 42 | 1,7645E+27 | 4,7964E+27 | 1,3038E+28 | 3,5441E+28 | 9,6339E+28 | 2,61876E+29 | 7,11853E+29 |
| 43 | 1,8065E+27 | 4,9106E+27 | 1,3348E+28 | 3,6285E+28 | 9,8633E+28 | 2,68111E+29 | 7,28802E+29 |
| 44 | 1,8485E+27 | 5,0248E+27 | 1,3659E+28 | 3,7129E+28 | 1,0093E+29 | 2,74347E+29 | 7,45751E+29 |
| 45 | 1,8905E+27 | 5,139E+27 | 1,3969E+28 | 3,7973E+28 | 1,0322E+29 | 2,80582E+29 | 7,627E+29 |
| 46 | 1,9326E+27 | 5,2532E+27 | 1,428E+28 | 3,8816E+28 | 1,0551E+29 | 2,86817E+29 | 7,79649E+29 |
| 47 | 1,9746E+27 | 5,3674E+27 | 1,459E+28 | 3,966E+28 | 1,0781E+29 | 2,93052E+29 | 7,96598E+29 |
| 48 | 2,0166E+27 | 5,4816E+27 | 1,4901E+28 | 4,0504E+28 | 1,101E+29 | 2,99287E+29 | 8,13547E+29 |
| 49 | 2,0586E+27 | 5,5958E+27 | 1,5211E+28 | 4,1348E+28 | 1,124E+29 | 3,05522E+29 | 8,30496E+29 |
| 50 | 2,1006E+27 | 5,71E+27 | 1,5521E+28 | 4,2192E+28 | 1,1469E+29 | 3,11757E+29 | 8,47445E+29 |
| 51 | 2,1426E+27 | 5,8242E+27 | 1,5832E+28 | 4,3036E+28 | 1,1698E+29 | 3,17993E+29 | 8,64394E+29 |
| 52 | 2,1846E+27 | 5,9384E+27 | 1,6142E+28 | 4,3879E+28 | 1,1928E+29 | 3,24228E+29 | 8,81342E+29 |
| 53 | 2,2266E+27 | 6,0526E+27 | 1,6453E+28 | 4,4723E+28 | 1,2157E+29 | 3,30463E+29 | 8,98291E+29 |
| 54 | 2,2687E+27 | 6,1668E+27 | 1,6763E+28 | 4,5567E+28 | 1,2386E+29 | 3,36698E+29 | 9,1524E+29 |
| 55 | 2,3107E+27 | 6,281E+27 | 1,7074E+28 | 4,6411E+28 | 1,2616E+29 | 3,42933E+29 | 9,32189E+29 |
| 56 | 2,3527E+27 | 6,3952E+27 | 1,7384E+28 | 4,7255E+28 | 1,2845E+29 | 3,49168E+29 | 9,49138E+29 |
| 57 | 2,3947E+27 | 6,5094E+27 | 1,7694E+28 | 4,8099E+28 | 1,3075E+29 | 3,55403E+29 | 9,66087E+29 |
| 58 | 2,4367E+27 | 6,6236E+27 | 1,8005E+28 | 4,8942E+28 | 1,3304E+29 | 3,61639E+29 | 9,83036E+29 |
| 59 | 2,4787E+27 | 6,7378E+27 | 1,8315E+28 | 4,9786E+28 | 1,3533E+29 | 3,67874E+29 | 9,99985E+29 |
| 60 | 2,5207E+27 | 6,852E+27 | 1,8626E+28 | 5,063E+28 | 1,3763E+29 | 3,74109E+29 | 1,01693E+30 |
| 61 | 2,5627E+27 | 6,9662E+27 | 1,8936E+28 | 5,1474E+28 | 1,3992E+29 | 3,80344E+29 | 1,03388E+30 |
| 62 | 2,6048E+27 | 7,0804E+27 | 1,9247E+28 | 5,2318E+28 | 1,4221E+29 | 3,86579E+29 | 1,05083E+30 |
| 63 | 2,6468E+27 | 7,1946E+27 | 1,9557E+28 | 5,3162E+28 | 1,4451E+29 | 3,92814E+29 | 1,06778E+30 |
| 64 | 2,6888E+27 | 7,3088E+27 | 1,9868E+28 | 5,4005E+28 | 1,468E+29 | 3,9905E+29 | 1,08473E+30 |